\documentclass[aps,prx,twocolumn,superscriptaddress,showpacs,floatfix,longbibliography]{revtex4-1}
\usepackage{amsmath,amssymb,graphicx}

\usepackage[utf8]{inputenc}
\usepackage[T1]{fontenc}
\usepackage{xcolor}

\IfFileExists{newtxtext.sty}
   {\usepackage{newtxtext,newtxmath}}
   {\IfFileExists{stix.sty}
      {\usepackage{stix}}
      {\IfFileExists{mathptmx.sty}
      {\usepackage{mathptmx}}{} } }

\usepackage{textcomp}

\usepackage{bm}

\IfFileExists{siunitx.sty}{\usepackage{booktabs,siunitx}}{}

\pdfoutput=1
\usepackage{color}
\definecolor{LinkColor}{rgb}{0.256,0.439,0.588}
\usepackage{hyperref}
\hypersetup{
   pdfauthor={Xiao Yan Xu},
   pdftitle={paper},
   colorlinks=true,
   citecolor=LinkColor,
   linkcolor=LinkColor,
   urlcolor=LinkColor
}

\renewcommand{\vec}[1]{\mathbf{#1}}

\usepackage{pifont}

\begin{document}

\title{Monte Carlo Study of Lattice Compact Quantum Electrodynamics with Fermionic Matter: \\ the Parent State of Quantum Phases}

\author{Xiao Yan Xu}
\email{wanderxu@gmail.com} \affiliation{Department of Physics,
Hong Kong University of Science and Technology, Clear Water Bay,
Hong Kong, China}
\author{Yang Qi}
\email{qiyang@fudan.edu.cn} \affiliation{Center for Field Theory
and Particle Physics, Department of Physics, Fudan University,
Shanghai 200433, China} \affiliation{State Key Laboratory of
Surface Physics, Fudan University, Shanghai 200433, China}
\affiliation{Collaborative Innovation Center of Advanced
Microstructures, Nanjing 210093, China}
\author{Long Zhang}
\affiliation{Kavli Institute for Theoretical Sciences and CAS
Center for Excellence in Topological Quantum Computation,
University of Chinese Academy of Sciences,  Beijing  100190,
China}
\author{Fakher F. Assaad}
\affiliation{Institut f\"ur Theoretische Physik und Astrophysik,
Universit\"at W\"urzburg, 97074, W\"urzburg, Germany}
\author{Cenke Xu}
\affiliation{Department of Physics, University of California,
Santa Barbara, CA 93106, USA}
\author{Zi Yang Meng}
\email{zymeng@iphy.ac.cn} 
\affiliation{Beijing National Laboratory for Condensed Matter Physics and Institute of Physics, Chinese Academy of Sciences, Beijing 100190, China} 
\affiliation{Department of Physics, The University of Hong Kong, China}
\affiliation{CAS Center of Excellence in Topological Quantum  Computation and School of Physical Sciences, University  of  Chinese Academy of Sciences,  Beijing  100190, China}
\affiliation{Songshan Lake Materials Laboratory, Dongguan, Guangdong 523808, China}

\begin{abstract}

The interplay between lattice gauge theories and fermionic matter
accounts for fundamental physical phenomena ranging
from the deconfinement of quarks in particle physics to quantum
spin liquid with fractionalized anyons and emergent gauge
structures in condensed matter physics. However, except for
certain limits (for instance large number of
flavors of matter fields), analytical methods can provide few
concrete results. Here we show that the problem of compact $U(1)$
lattice gauge theory coupled to fermionic matter in $(2+1)$D is
possible to access via sign-problem-free quantum Monte Carlo
simulations.  One can hence map out the phase diagram as a function of fermion flavors
and the strength of gauge fluctuations. By
increasing the coupling constant of the gauge field, gauge
confinement in the form of various spontaneous symmetry breaking
phases such as valence bond solid (VBS) and N\'eel
antiferromagnet emerge. Deconfined phases with
algebraic spin and VBS  correlation functions  are also observed. Such
deconfined phases are incarnation of an exotic states of matter,
$i.e.$ the algebraic spin liquid, which is generally viewed as the
parent state of various quantum phases. The phase transitions
between  deconfined and confined phases, as well as that
between the different confined phases provide various
manifestations of  deconfined quantum criticality. In
particular,   for four flavors,  $N_f = 4$, our data suggests a continuous quantum
phase transition between the VBS and N\'{e}el order. We also
provide  preliminary theoretical  analysis for these quantum phase
transitions.

\end{abstract}

\date{\today}
%\date{July 20, 2018}

\maketitle

\section{Introduction}
\label{sec:introduction}

The interplay between lattice gauge theories and fermionic matter
has allured the imagination of physicists for several
decades~\cite{AffleckMarston1988,MarstonAffleck1989,Fradkin1990,Fiebig1990,Lee1998,Rantner2001,Rantner2002,Senthil2000,Senthil2002,Herbut2003,Hermele2004,
Assaad2005,Hermele2005,Hermele2007,Nogueira2008}. This is because
gauge fields coupled to  matter fields is a fundamental concept in
many areas of physics. For example, in condensed matter, $(2+1)$D
field theories with a compact $U(1)$ gauge field coupled to
gapless relativistic fermions often serve as the low energy
effective field theories in 2D strongly correlated electron
systems including cuprate
superconductors~\cite{AffleckMarston1988,MarstonAffleck1989,Lee1998,Rantner2001,Rantner2002}
and quantum spin
liquids~\cite{Hermele2004,Assaad2005,Hermele2005,Hermele2007,He2017}.
In high energy physics, the mechanism of quark confinement in
gauge theories with dynamical fermions such as quantum
chromodynamics (QCD) is among the most elusive subjects, and the
absence/presence of a deconfined phase in 3D compact quantum
electrodynamics (cQED$_3$) coupled to (not necessarily large)
$N_f$ massless fermions has attracted lots of
attention~\cite{Fiebig1990,Herbut2003,Hermele2004,Assaad2005,Hermele2005,Hermele2007,Nogueira2008},
and remains unsolved to this day.

In recent years, collective efforts from both condensed matter and
high energy physics have started to generate promising
outcomes~\cite{Fiebig1990,Assaad2005,Armour2011,Assaad2016,Gazit2017,Gazit2018,ncQED1,ncQED2,ncQED3,QED1,QED2,Prosko2017}.
There exist concrete examples, by now, of discrete $Z_2$ gauge
field theories coupled to fermionic matter at $(2+1)$D, deconfined
phase with fractionalized fermionic excitations at weak gauge
fluctuation, as well as symmetry breaking phase with gapped
fermionic excitations at strong gauge fluctuation have been
observed~\cite{Assaad2016,Gazit2017,Gazit2018}. The apparently
continuous transition between deconfined and confined phases is
highly non-trivial~\cite{Gazit2018}  as it is driven by the
condensation of emergent fractionalized excitations and is hence
beyond the scope of  the Landau-Ginzburg-Wilson paradigm of
critical phenomena in which symmetry-breaking is described by a
local order parameter.

\begin{figure*}[htp!]
\includegraphics[width=\textwidth]{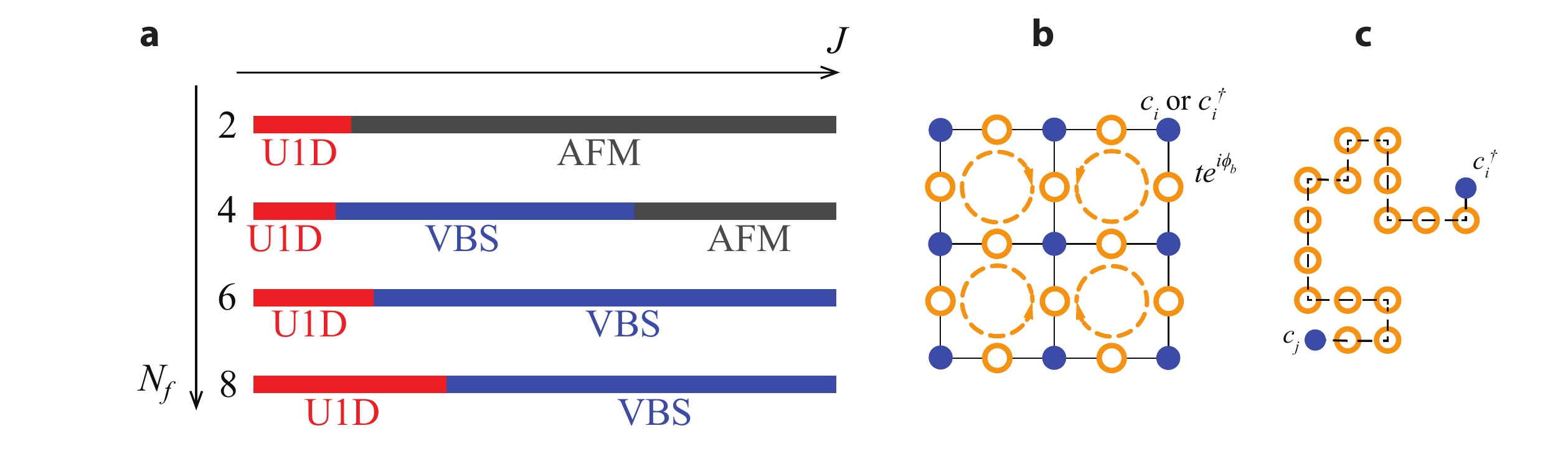}
\caption{\textbf{a}. Phase diagram spanned by the fermi flavors
$N_f$ and the strength of gauge field fluctuations $J$ of the
model shown in \textbf{b}. U1D stands for the $U(1)$ deconfined
phase where the fermions
dynamically form a Dirac system.  This phase  corresponds to the
algebraic spin liquid where all correlation functions show slow
power-law decay. VBS stands for valence bond solid phase and AFM
stands for the antiferromagnetic long-range ordered phase (N\'eel
phase).
\textbf{b}. Sketch of
the model of Eq.~\eqref{eq:hamiltonian}. The yellow circles
represent  the gauge field attached to each fermion hopping and
the yellow dashed lines stand for the flux term per plaquette.
\textbf{c}. The gauge invariant propagator for fermions with
string of gauge fields attached.} \label{fig:phasediagram}
\end{figure*}

The cQED$_3$ is the simplest theory to discuss confinement
and chiral symmetry breaking~\cite{Polyakov1977,Banks1977,Fosco2006}.
The pure cQED$_3$ without matter field is known to be always
confining~\cite{Mandelstam1976,Polyakov1977,Herbut2003,Case2004}.
However, when there is fermionic matter, the gapless fermionic
fluctuations may drive the system towards deconfinement. The
large $N_f$ limit of cQED$_3$ with fermionic matter is believed
to belong to this case~\cite{Hermele2004,Hermele2005,Unsal2008,Nogueira2008},
but the existence of deconfined phase for small $N_f$ is still under debate~\cite{Kragset2004,Assaad2005,Borkje2005,Fiore2005,Armour2011,pufu2014}.
Analytically, the perturbative calculation at small $N_f$ is uncontrolled.
Numerically, recent Hybrid Monte Carlo (HMC) calculations in~\cite{Armour2011} face the difficulties caused by fermion zero modes. Though these difficulties may be cured by turning on a four-fermion interaction term, the scaling dimension of the four fermion interaction will receive corrections from gauge fluctuation at the order of $1/N_f$, which may lead to a relevant run away flow~\cite{alicea2005}. Thus a combined RG flow of monopole and four fermion interactions may be complicated, and a deconfined phase could still exist in the phase diagram but evades the previous study of HMC.
cQED$_3$ with finite $N_f$ flavors of fermionic matter is
particularly important to condensed matter physics because these
cases actually correspond to the low energy field theory
description of many interesting strongly correlated electron
systems, and therefore host the potential promise of establishing
the new paradigms in condensed matter physics. And furthermore,
perturbative renormalization group calculation to higher orders
have recently been carried out in attempt to accquire the critical
properties of the deconfinement to confinement transition in form
of QED$_3$-Gross-Neveu universality
classes~\cite{Janssen2017,Ihrig2018,Nikolai2018,Gracey2018}.

Based on these considerations, in this work, we succeeded in
performing large-scale quantum Monte Carlo (QMC) simulations on
the cQED$_3$ coupled to $N_f$-flavor of fermions, and eventually
map out the phase diagram (Fig.~\ref{fig:phasediagram})  in the
 fermion flavor and gauge field
fluctuations strength plane. Deconfined phases -- $U(1)$ deconfined phase (U1D
hereafter) to be more precise -- are indeed found in the phase
diagram for $N_f=6$ and 8, and even at $N_f=2$ and $4$ there are
very positive signatures of their existence. Various confined
phases, in the form of different symmetry breakings, such as
antiferromagnetic order (AFM) and valence bond solid (VBS), are
also discovered. Interesting quantum phase transitions, between
deconfined and confined phases, and between different confined
phases
\cite{Senthil2004,Sandvik2007,Assaad2016,Qin2017,Sato2017,You2018}
are revealed as well.

For the sake of smoother narrative, the rest of the paper is
organized in the following order. In Sec.~\ref{sec:model} and
\ref{sec:symmetrylimit}, we first start with a quantum rotor model
coupled to fermions, which can be formulated as cQED$_3$ coupled
to fermionic matter. Then in Sec.~\ref{sec:sign}, we discuss
the sign structure of this model, where we find  that a pseudo-unitary
group can be used to avoid the phase problem at  odd $N_f$, and
the sign problem  at even  $N_f$. In
Secs.~\ref{sec:difficulty} and ~\ref{sec:fast}, we explain the
challenges in the QMC simulation even without sign-problem and
provide our solution with a fast update algorithm for simulating
gauge fields with continuous symmetries. In Sec.~\ref{sec:results} we
discuss the whole phase diagram, and then focus on the physical
properties and understanding of the U1D phase, in particular the
reason for it being the parent state of various quantum phases, the
deconfinement-confinement phase transitions, and VBS to AFM phase
transition at  $N_f=4$. Preliminary theoretical analysis
of these transitions  is also given in Sec.~\ref{sec:results}.
Finally, the discussion and conclusions are given in
Sec.~\ref{sec:discuss}.

\section{Model and Method}
\subsection{Rotor model with fermion}
\label{sec:model}

The system we are interested in, can be most conveniently
formulated as a 2D quantum rotor model coupled to fermions with
Hamiltonian
\begin{eqnarray}
H&=&\frac{1}{2}JN_{f}\sum_{\langle i,j \rangle} \frac 1 4 \hat{L}^{2}_{ij}-t\sum_{\langle i,j \rangle\alpha}\left(\hat{c}^{\dagger}_{i\alpha}e^{i\hat{\theta}_{ij}}\hat{c}_{j\alpha}+\text{h.c.}\right) \nonumber\\
&& +\ \frac{1}{2}K\ N_f\sum_{\square}\cos \left( \text{curl} \hat{\theta} \right),
\label{eq:hamiltonian}
\end{eqnarray}
where $\hat{L}_{ij}$ and $\hat{\theta}_{ij}$ are canonical angular
momentum, $[\hat{L}_{ij},e^{\pm i \hat{\theta}_{ij}}] =
\pm e^{\pm i\hat{\theta}_{ij}}$,  and its coordinate operator of rotors on each bond
$b=\langle ij \rangle$ of a 2D square lattice, as depicted in
Fig.~\ref{fig:phasediagram} (b). The fermion
flavor $\alpha$ runs from 1 to $N_{f}$ and the fermions are
minimally coupled to the rotor via nearest-neighbor hopping on the
square lattice. The flux term with $K>0$ favors $\pi$-flux  in
each elementary plaquette $\square$, where the magnetic flux of
each plaquette $\square$ is defined as $\text{curl} \hat{\theta} =
\sum_{b\in\square} \hat{\theta}_{b}$ and the summation over
$\hat{\theta}_b$ has been taken in either clockwise or
anticlockwise  orientation around  an elementary plaquette.

For the Monte Carlo simulations, it is convenient to work  in a
representation where  $\hat{\theta}_{ij}$ is diagonal. That is,
omitting the bond index, $\hat{\theta} | \phi \rangle  = \phi  |
\phi \rangle  $     with  $ \phi \in \left[ 0, 2 \pi  \right) $.
In this representation, $\hat{L} = -i \frac{\partial}{\partial
\phi} $    with eigenvectors  $\hat{L} \langle \phi | l \rangle
\equiv  \hat{L} e^{i \phi l }  = l  e^{i \phi l } $ and $l \in
\mathbb{Z}$. With these  definitions, the  resolution of unity
reads:  $ \int_{0}^{2\pi}   | \phi \rangle  \langle \phi | =
\frac{1}{2\pi }\sum_{l} | l \rangle \langle l | = \hat{1} $. To
formulate the path integral, we have to estimate  the matrix
element:   $  \langle \phi' | e^{ - \Delta \tau  J N_f \hat{L}^2
/8} | \phi \rangle $. To this end we insert resolution of the unit
operator, and  use the   Poisson summation  formula to obtain:
\begin{equation}
  \langle \phi' | e^{ - \Delta \tau  J N_f \hat{L}^2 /8} | \phi \rangle
\sim
 e^{  -\frac{4}  {\Delta \tau  J N_f }   \left( 1 - \cos( \phi - \phi' )\right)  },
\end{equation}
where the Villain approximation is used. With
the  above, the Hamiltonian in Eq.~\eqref{eq:hamiltonian} can be
formulated in a coherent-state path integral with action
$S=S_F+S_{\phi}=\int_0^{\beta}d\tau (L_F+L_\phi)$ and the
Lagrangian for fermion and gauge field parts are
\begin{eqnarray}
L_F &=& \sum_{\langle ij \rangle\alpha}{\psi}^{\dagger}_{i\alpha} \left[(\partial_\tau -\mu)\delta_{ij}-t e^{i\phi_{ij}}   \right]   {\psi}_{j\alpha} + \text{h.c.}, \nonumber \\
L_\phi &=& \frac{4} {JN_{f}\Delta \tau ^2} \sum_{\langle ij \rangle}
\left( 1-\cos(\phi_{ij}(\tau+1)-\phi_{ij}(\tau)) \right) \nonumber \\
&& +\frac{1}{2}K N_f\sum_{\square}\cos \left( \text{curl} \phi \right),
\label{eq:lagrangian}
\end{eqnarray}
respectively, where $\mu$ will be set to zero for half-filled
case. $\beta=\frac 1 T$ is the inverse temperature.  The model in
Eq.~\eqref{eq:hamiltonian} has now  been explicitly formulated as
(unconstrained) cQED$_3$ coupled to fermionic
matter~\cite{Hermele2004,Hermele2005,Armour2011}. We will now
consider the symmetries of the model and show that the Gauss law
is dynamically imposed in the low temperature
limit.

\subsection{Symmetries and limiting cases}
\label{sec:symmetrylimit}
Our model, see Eq.~\eqref{eq:hamiltonian}, has global and local
symmetries. It enjoys a manifest global $SU(N_f)$  spin symmetry
as well as a  particle hole-symmetry:
\begin{equation}
\hat{P}^{-1}   z \hat{c}^{\dagger}_{i\alpha} \hat{P} =
\overline{z} (-1)^{i}    \hat{c}^{\phantom\dagger}_{i\alpha}.
\end{equation}
In the above, $z$ is a complex number  that  makes it clear that
the particle-hole symmetry is anti-unitary, and $(-1)^{i}$ takes
the value $1$ ($-1$) on sublattice A (B).

The local $U(1)$ gauge  transformation
\begin{equation}
\hat{c}_{i \alpha} \to \hat{c}_{i \alpha}e^{i\varphi_i}, \;\; \hat{\theta}_{ij}\to\hat{\theta}_{ij}+\varphi_i - \varphi_j,
\end{equation}
is an invariant. The generator of this local symmetry corresponds
to a local conserved charge (Gauss law)
\begin{equation}
\hat{Q}_{i} = -\sum_{j}\hat{L}_{ij} + \sum_{\alpha} \left( \hat{c}^{\dagger}_{i\alpha}\hat{c}^{\phantom\dagger}_{i\alpha} - 1/2 \right)
\label{eq:Qi}
\end{equation}
with $[\hat{Q}_{i},\hat{H}]=0$. In our simulations we sample over
all $\hat{Q}_{i}$ sector, such that our Hamiltonian corresponds to
an {\it unconstrained} gauge theory. As a consequence, correlation
functions of gauge  dependent quantities such as the single
particle operator are local in space but not it  time: $ \langle
\hat{c}^{\dagger}_{i\alpha}(\tau) \hat{c}_{j\alpha} \rangle = \delta_{i,j} \langle \hat{c}^{\dagger}_{i\alpha}(\tau) \hat{c}_{i\alpha} \rangle
$.  Below we argue that the Gauss law constraint is dynamically
imposed in the zero temperature limit.

At $ J  = \infty $,  $\hat{L}_{ij}$ vanishes and charges are
completely localized since hopping on a given bond involves
excitations of the rotor mode. In this limit charge
configurations, corresponding to specific values of $\hat{Q}_{i}
$, are degenerate and at any finite value of $J$  the degeneracy
will be lifted. The dynamical generation of the term
\begin{equation}
\hat{H}_{Q}  = \sum_{i,j}  K_{i,j}   \hat{Q}_i \hat{Q}_j  +
\cdots.
\end{equation}
accounts for the lifting of this degeneracy. Note that since
$\left\{ \hat{P}, \hat{Q}_i \right\} = 0$ terms containing
products of odd numbers of $\hat{Q}_i$'s are forbidden. The above
equation defines a {\it classical} model with a finite temperature
Kosterlitz-Thouless transition. At zero temperature the $\hat{Q}_i
$ are frozen in a given pattern, and the Gauss law is imposed.

For $J \rightarrow \infty $ our model maps onto an  $SU(N_f)$ quantum
antiferromagnetic. We again start from the $J = \infty  $
degenerate case, and consider $t$ in second order degenerate
perturbation theory. As mentioned above, hopping of a fermion with
flavor  index $\alpha$ from site i to nearest neighbor site j leaves
the rotor in an excited state, associated with energy cost $J$.
The only way to remove this excitation is  for a fermion with
flavor index $\alpha'$ to hop back  from site j to site i. These
processes are encoded in the $SU(N_f)$ Heisenberg model:
\begin{equation}
\hat{H}_{J\rightarrow \infty } \propto - \frac{t^2}{J}
\sum_{\langle i j \rangle }  \left( \hat{D}_{ij}^{\dagger}
\hat{D}_{ij}^{\phantom\dagger}  +
\hat{D}_{ij}^{\phantom\dagger}   \hat{D}_{ij}^{\dagger} \right)
\end{equation}
with $\hat{D}_{ij} = \sum_{\alpha= 1}^{N_f}  \hat{c}^{\dagger}_{i\alpha}   \hat{c}^{\phantom\dagger}_{j\alpha}   $.
In our simulations we have on average $N_f/2$ fermions per site
such that the representation of the $SU(N_f)$ group corresponds to
the antisymmetric self-adjoint representation (i.e. Young tableau
corresponding to a column of $N_f/2$ boxes) \cite{sachdevread}.  On the square lattice
and for even values of $N_f$ where the negative sign problem is
absent, this model has been considered in former auxiliary field
QMC simulations \cite{Assaad2005}. At $N_f = 2$ one finds an
antiferromagnetic state  and  at and beyond $N_f = 6$ a  VBS
state.  In the large-$N_f$ limit we recover the Marston-Affleck
\cite{MarstonAffleck1989} saddle point accounting for
dimerization. At $N_f =4$ and in the absence of charge
fluctuations, Ref.~\cite{Assaad2005}  finds no compelling evidence
of VBS and AFM orders when considering lattices up to $24 \times
24$. On the other hand, simulations of the corresponding $SU(4) $
Hubbard model~\cite{Wang14} are consistent with an AFM state in
the large $U$ limit albeit with decreasing value of the order
parameter as a function of $U$. In our simulations charge
fluctuations are present and the phase diagram is consistent with
AFM order in the large $J$ limit.

At $J=0$, $\hat{\theta}_{ij}$ becomes a classical variable in the sense that it has no imaginary time dynamics.    Even
in the absence of the flux term, the coupling to  the fermions
favors,  according to  Lieb's theorem \cite{Lieb1994}, a
$\pi$-flux per  plaquette, with associated dynamically generated
Dirac dispersion relation of the $\hat{c}$-fermions. The fate of
this state at low values of $N_f$  and  when gauge fluctuations
are accounted for is  one of the central  aims of our research.

\subsection{Absence of sign problem for even $N_f$}
\label{sec:sign}

To simulate above model with quantum Monte Carlo method, we start
with the partition function
\begin{equation}
Z=\int D(\phi,\bar{\psi}, \psi) e^{-(S_\phi +S_F)} = \int D\phi e^{-S_\phi} \text{Tr}_{\psi} \left[ e^{-S_F}  \right].
\end{equation}
As the action of gauge field part $S_{\phi}=\int_0^{\beta}d\tau \;
L_\phi$ with $L_{\phi}$ shown in Eq.~\eqref{eq:lagrangian}, is
always real (thus  its exponential is always positive), the sign
structure of the Monte Carlo weight will only come from the trace
over fermions. To trace out fermion, we first discretize the
imaginary time $\tau= z \Delta\tau$ ($z=1,2,\cdots, L_\tau$) where
$L_{\tau}$ is the total number of time slices ($L_{\tau}
\Delta\tau=\beta$), then performing the fermion trace, we have
\begin{equation}
\text{Tr}_{\psi}\left[ e^{-S_F}  \right]  = \left[ \det\left(I + \prod_{z=1}^{L_{\tau}} {B_z} \right) \right] ^{N_f},
\label{eq:determinant}
\end{equation}
with $S_{F}=\int_0^{\beta} d\tau \; L_{F}$ and $L_{F}$ shown in
Eq.~\eqref{eq:lagrangian}, and after the discretization of
$\beta$, $B_z=e^{-\Delta\tau V_z }$ with $V_z$ the coupling matrix
for each fermion flavor (we have $N_f$ in total, therefore there
is power $N_f$ in Eq.~\eqref{eq:determinant}), which only has
elements connecting sites between different sublattice of the
square lattice, $(V_z)_{ij}=-te^{i\phi_{ij}}$. We recognize that such kinds of $B_z$ matrices
form a pseudo-unitary group $SU(n,n)$ where $2n$ is dimension of
$B_z$ (total number of sites). As proved in
Appendix~\ref{sec:psu},  $\forall D \in SU(n,n)$,  $\det(1+D)\in
\Re$ holds. Therefore the fermion weight will always be real for
all integer $N_f$ and, most importantly, be semipositive-definite
for all even value of $N_f$ such that QMC simulations can be
performed. Although the current paper only focus on the compact
$U(1)$ gauge fields, the $SU(n,n)$ group actually allows one to
add extra non-Hermitian terms to the model, such as a staggered
imaginary chemical potential,  that are  also sign problem free
for even values of $N_f$. It will be very interesting to study the
non-Hermitian models and their properties, in the presence of
gauge field fluctuations and electronic interactions in future
investigations.

\subsection{Difficulties of the QMC simulation}
\label{sec:difficulty}

Although there is no sign-problem for even $N_f$, the simulation
of the Eq.~\eqref{eq:hamiltonian} is by no means simple. Earlier
attempts in the high energy community have been devoted to
simulate similar models by means of hybird Monte
Carlo~\cite{DUANE1987,Fiebig1990,Fiore2005,Armour2011} with
(dynamical) mass term.  Mass terms are essential to avoid
divergences when calculating forces in the realm of
Hamiltonian~\cite{Beyl2018}  and Langevin~\cite{BATROUNI1987}
dynamics.  Mass terms  however introduce biases the severity of
which have to be a posteriori clarified.

On the other hand, in the condensed matter community, the
determinantal QMC (DQMC) is more popular and it usually uses local
updates and the mass terms  are not  essential
here~\cite{Blankenbecler1981,White1989,Assaad2008,Shi2016}.
However, as far as we are aware of, there exists no DQMC
simulation on cQED$_3$ coupled to fermionic matter, and our work
hence serves as a first attempt. As explained below, to be able
to simulate the model in Eq.~\eqref{eq:hamiltonian}, there are
several obstacles one needs to overcome.

The most obvious obstacle is about the computational complexity.
For general models the computation complexity of DQMC for one
sweep is $\mathcal{O}(\beta N^4)$ (here $N$ is total number of
sites), for models where fast update is applicable the complexity
can be reduced to $\mathcal{O}(\beta N^3)$. The most common
example is Hubbard model with onsite interaction, when we flip a
single auxiliary field at time slice $z$ and site $i$, it only
changes one diagonal value in $B_z$ matrix, then the new equal
time Green's function $G'(\tau,\tau)$ can be calculated as

\begin{align}
G' & =(I+B_{1}B_{2}\cdots B_{z-1}(I+\Delta)B_{z}\cdots B_{L_{\tau}-1}B_{L_{\tau}})^{-1}\nonumber \\
 & =G[I+\Delta(I-G)]^{-1}
 \label{eq:newg}
\end{align}
where $G$ is the Green's function at previous step. As
$\Delta_{ii}$ is the only non-zero element in $N\times N$ matrix
$\Delta$, $\Delta(1-G)$ can be written as the outer product of two
vectors, then Sherman-Morrison formula can be used to reduce the
complexity for calculating Eq.~\eqref{eq:newg} from
$\mathcal{O}(N^3)$ to
$\mathcal{O}(N^2)$~\cite{Blankenbecler1981,White1989}. Such update
scheme is referred as \emph{fast update}. For one sweep over all
auxiliary fields (scales as $\mathcal{O}(\beta N)$), the scaling
will be $\mathcal{O}(\beta N^3$) instead of $\mathcal{O}(\beta
N^4)$. For models with off-site interaction,  for example in our
case due to the coupling between fermion and gauge field, usually
we need to make a further Suzuki-Trotter decomposition over all
bonds (assume bond $b$ connects site $i$ and $j$, total number of
bonds is $N_b$), and $B_z$ matrix will be written as

\begin{equation}
B_z = \prod_{b=1}^{N_{b}} B_{z,b},
\end{equation}
where $B_{z,b}=e^{h_{z,b}}$ and $h_{z,b}$ is a $N\times N$ matrix
with only non-zero elements  $B_{z,b}(i,j)=B_{z,b}(j,i)^*=
f(\phi_{ij})$ where $f(\phi_{ij})=\rho(\phi_{ij})
e^{i\theta(\phi_{ij})}$ is the complex function of auxiliary field
$\phi_{ij}$ in polar form. It is obvious that only when
$\theta(\phi_{ij})=0$ or $\pi$, $B_{z,b}(i,j)$ will be
diagonalized by an auxiliary field independent unitary
transformation, and then the aforementioned fast update scheme
applies. There are many known models belong to this case, such as
model with Heisenberg type
interaction~\cite{He2016,He2016SMG,Wu2016}, models with $Z_2$
(bosonic) gauge field coupled to fermionic
matter~\cite{Xu2017,Assaad2016,Gazit2017,YYHe2018}, etc. Novel
physics has been found in these models, for example in
Ref.~\cite{He2016SMG} a continuous phase transition with fermion
mass generation without spontaneously breaking any symmetry was
identified (similar physics was also found in the lattice QCD
community~\cite{shailesh1,shailesh2,shailesh3,Catterall2016}).
Unfortunately, our model in Eq.~\eqref{eq:hamiltonian} involves
$U(1)$ gauge fields, the auxiliary field $\phi_{ij}$ is therefore
a continuous variable and $\theta(\phi_{ij})=\phi_{ij}$, thus our
model does {\it not} belong to the cases discussed above and
naively the fast update cannot be applied.

\subsection{Fast update algorithm designed for $U(1)$ gauge fields}
\label{sec:fast}

There is indeed  an alternative way to to design  a fast update
algorithm for the  model in Eq.~\eqref{eq:hamiltonian}. It is
based  on the Woodbury matrix identity  that effectively
generalizes the Sherman-Morrison formula  to higher rank matrices.
We recognize that the  new $B'_{z,b}$ after a single update can
directly be factorized into $(1+\Delta) B_{z,b}$ with
\begin{widetext}
\begin{equation}
\left[\begin{array}{cc}
\Delta_{ii} & \Delta_{ij}\\
\Delta_{ji} & \Delta_{jj}
\end{array}\right]=\left[\begin{array}{cc}
\left(1-e^{-i(\phi_{ij}-\phi_{ij}')}\right)\sinh^{2}\Delta\tau & \left(-e^{i\phi_{ij}}+e^{i\phi_{ij}'}\right)\sinh\Delta\tau\cosh\Delta\tau\\
\left(-e^{-i\phi_{ij}}+e^{-i\phi_{ij}'}\right)\sinh\Delta\tau\cosh\Delta\tau & \left(1-e^{i(\phi_{ij}-\phi_{ij}')}\right)\sinh^{2}\Delta\tau
\end{array}\right]
\end{equation}
\end{widetext}
and other elements of the $N\times N$ matrix $\Delta$ are zero.
With such kind of structure and note that Eq.~\eqref{eq:newg}
still holds, $\Delta(1-G)$ can be written as product of two
matrices with dimension $N\times 2$ and $2\times N$. Therefore, we
can use the generalized version of Sherman-Morrison formula
(Woodbury matrix identity) to calculate Eq.~\eqref{eq:newg}, which
also has complexity $\mathcal{O}(N^2)$.

With such special designed fast update, we are now ready to
simulate cQED$_3$ coupled to fermionic matter without artificial
mass term and still enjoy the $\mathcal{O}(\beta N^3)$
computational complexity.

\section{Results}
\label{sec:results}

\subsection{Physical observables}
\label{sec:observables}

Our model and major results are schematically summarized in
Fig.~\ref{fig:phasediagram} (a) and (b), respectively, but before
starting the discussion of QMC results, we first introduce the QMC
observables that are used to characterize symmetric and symmetry
breaking phases and their phase transitions. Since physical observables are Hermitian, we constructed
and measured various gauge invariant structure factors, including
spin $\chi_S(\vec{k})$ and dimer $\chi_D(\vec{k})$ structure
factors. They are defined as
\begin{equation}
\chi_{S}(\vec{k})=\frac{1}{L^{4}}\sum_{ij}\sum_{\alpha\beta}\langle
S_{\beta}^{\alpha}(i)S_{\alpha}^{\beta}(j)\rangle
e^{i\vec{k}\cdot(\vec{r}_{i}-\vec{r}_{j})}
\end{equation}
\begin{equation}
\chi_{D}(\vec{k})=\frac{1}{L^{4}}\sum_{ij}\left(\langle
D_{i}D_{j}\rangle-\langle D_{i}\rangle\langle
D_{j}\rangle\right)e^{i\vec{k}\cdot(\vec{r}_{i}-\vec{r}_{j})}
\end{equation}
where the spin operator
$S_{\beta}^{\alpha}(i)=c_{i\alpha}^{\dagger}c_{i\beta}-\frac{1}{N_{f}}\delta_{\alpha\beta}\sum_{\gamma}c_{i\gamma}^{\dagger}c_{i\gamma}$,
and the dimer operator
$D_{i}=\sum_{\alpha\beta}S_{\beta}^{\alpha}(i)S_{\alpha}^{\beta}(i+\hat{x})$
is defined as dimer along the nearest-neighbor bond in $\hat{x}$
direction.

From these structure factors, one can further construct
dimensionless quantities -- the correlation
ratio~\cite{TCLang2016} -- to determine the precise position of phase transitions. If one would like to detect the
transition towards antiferromagnetic long range order, the
antiferromagnetic correlation ratio is
\begin{equation}
r_{\text{AFM}}=1-\frac{\chi_{S}(\vec{X}+\delta\vec{q})}{\chi_{S}(\vec{X})},
\end{equation}
where $\vec{X}=(\pi,\pi)$ is the order wavevector for AFM on the
square lattice and $\delta\vec{q}=(\frac{2\pi}{L},0)$ is the
smallest momentum away from $\vec{X}$. In the same vein, we define
correlation ratio for the valence bond solid (VBS) order from
dimer structure factor
\begin{equation}
r_{\text{VBS}}=1-\frac{\chi_{D}(\vec{M}+\delta\vec{q})}{\chi_{D}(\vec{M})},
\end{equation}
where $\vec{M}=(\pi,0)$ is the order wavevector for VBS. Other
quantities, such as the energy density and various correlation
functions (spin, dimer) in real space, are also measured in the
QMC simulation.

\subsection{Phase diagram}
\label{sec:phasediagram}

Now we can discuss the results from QMC simulation of
Eq.~\eqref{eq:hamiltonian}. Starting with the final phase diagram
that schematically summarized all the data, as shown in
Fig.~\ref{fig:phasediagram} (a), the phase diagram is spanned
along the axes of $N_f$ and $J$. We set $K=t=1$ as the
energy unit and choose the gauge fluctuation strength $J$ as the
tuning parameter to study different $N_f$ cases. For each $N_f$
there are different phases and phase transitions, but there are
similarities for all $N_f$ investigated, that is, at small $J$,
$U(1)$ deconfined phases (U1D in Fig.~\ref{fig:phasediagram} (a))
are universally present in the phase diagram. This finding is
highly non-trivial, as explained in the introduction
(Sec.~\ref{sec:introduction}), from both high energy physics and
condensed matter physics communities, the existence of such a
deconfined phase in cQED$_3$ is still under debate, due to the
lack of controlled calculation at finite and small
$N_f$~\cite{Fiebig1990,Lee1998,Rantner2001,Rantner2002,Senthil2000,Senthil2002,Herbut2003,Hermele2004,
Assaad2005,Hermele2005,Hermele2007, Nogueira2008}, and our finding
presented here provide the first set of concrete evidence to
support the existence of this phase.

Moreover, as will be further elucidated in later sections, such a
deconfined phase is expected to be the algebraic spin
liquid~\cite{Rantner2001,Rantner2002,Senthil2000,Senthil2002,Hermele2005,Hermele2007,Xu2007},
in which critical correlations of many competing order parameters, such as antiferromagnetic order,
valence bond solid order, charge density wave and
superconductivity, coexist and share the same
power-law decay due to the $U(1)$
gauge deconfinement and the subsequential conformally invariant,
interacting fixed point~\cite{Hermele2005,Hermele2007}. Starting
from the algebraic spin liquid phase, one could easily apply
various perturbations and drive the system into various
symmetry-breaking phases. Therefore, the algebraic spin liquid
phase, i.e. U1D discovered in this work, actually serves as the
original state of many interesting quantum phases, hence dubbed
{\it parent state of quantum phases}. The discovery of such a
deconfined phase, is the most important result of this work.

As $J$ increases, the system goes through deconfine-confine phase
transitions to various symmetry-breaking phases. In the case of
$N_f=2$, the symmetry-breaking phase is the AFM (N\'eel) phase,
whereas in the case of $N_f=4$, the symmetry-breaking phases are
VBS and AFM phases. And further increases $N_f$, the
symmetry-breaking phases are VBS solely. According to
Ref.~\cite{Gazit2018}, the deconfine-confine phase transition
could be a version of the deconfined quantum critical point with
emergent continuous symmetry, and besides the QMC data, we will
also provide a preliminary field theoretical description of this
transition. The transition from VBS to AFM phases inside the
confined regime at $N_f=4$, if it is indeed continuous as our data
suggest, is also a deconfined quantum critical
point~\cite{Senthil2004,Sandvik2007,Nahum2015,Assaad2016,Qin2017,Sato2017,You2018,
Ma2018}, whose theory will be explained later.

Overall, the presence of the U1D deconfined phase, and the phase
transitions between deconfined to confined phases and within the
confined phases, are all intriguing and show the rich physics
behind the simple model of Eq.~\eqref{eq:hamiltonian}.

Below, we discuss some exotic aspects of the phases and phase
transitions we obtain.

\begin{figure}[t!]
\includegraphics[width=\columnwidth]{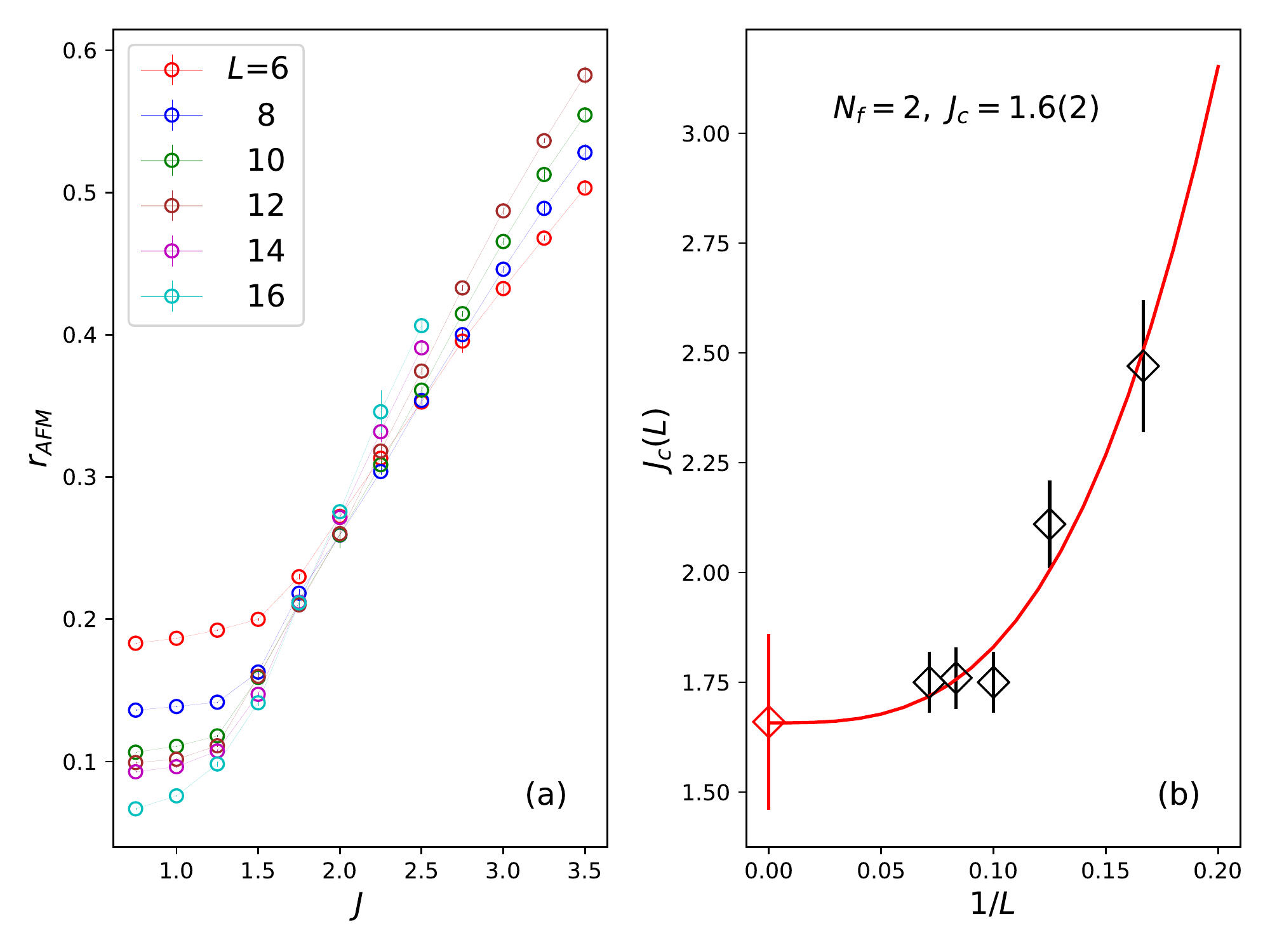}
\caption{(a) The antiferromagnetic correlation ratio through U1D to AFM transition at $N_f=2$. Here $\beta=4L$,  $\Delta \tau=0.2$. The crossing points are the transition point separating the deconfined phase and N\'eel phase. (b) The $1/L$ extrapolation of the crossings estimates the U1D to AFM transition point $J_c=1.6(2)$ for $N_f=2$.}
\label{fig:n2rafm}
\end{figure}

\subsection{U1D phase and confinement transition}
\label{sec:phaseboundaries}
\subsubsection{$N_f=2$}
First we focus on the case $N_f=2$. In the  static limit
($J\rightarrow 0$), the  gauge fields   are frozen into a
$\pi$-flux  pattern per plaquette~\cite{Lieb1994,Tanaka2005}  that
results in a  Dirac gapless dispersion relation.  At  finite $J$,
the  gauge fields fluctuate  and   proliferation of  monopoles of
the gauge fields may drive spinons (the fermions) to confine. On
the other hand, at large values of  $J$, an $SU(N_f)$
antiferromagnetic effective low energy model in the self-adjoint
antisymmetric representation  emerges (see
Sec.~\ref{sec:symmetrylimit}).  At $N_f =2$ the ground state of
this model is known to  host an AFM.

Fig.~\ref{fig:n2rafm}(a) shows the AFM correlation ratio
$r_{\text{AFM}}$ as a function of $J$ for different system sizes. Fig.~\ref{fig:n2rafm}(b) shows the crossing points of pairs of adjacent system sizes and a power-law extrapolation in $1/L$ gives rise to an estimate of the
confinement  transition in the thermodynamic limit:  $J_c =
1.6(2)$. Since the correlation ratio shows no abrupt  features,
the transition from the deconfined phase to the AFM  is more
likely to be continuous. This is consistent with the flux energy
per plaquette data presented in Fig.~\ref{fig:n2fluxenergy} in
Appendix~\ref{sec:fluxenergy}.  For the same  $J$  values as
considered in Fig.~\ref{fig:n2rafm}  the flux energy per plaquette
does not develop a sharp change of slopes.

\begin{figure}[t!]
\includegraphics[width=0.9\columnwidth]{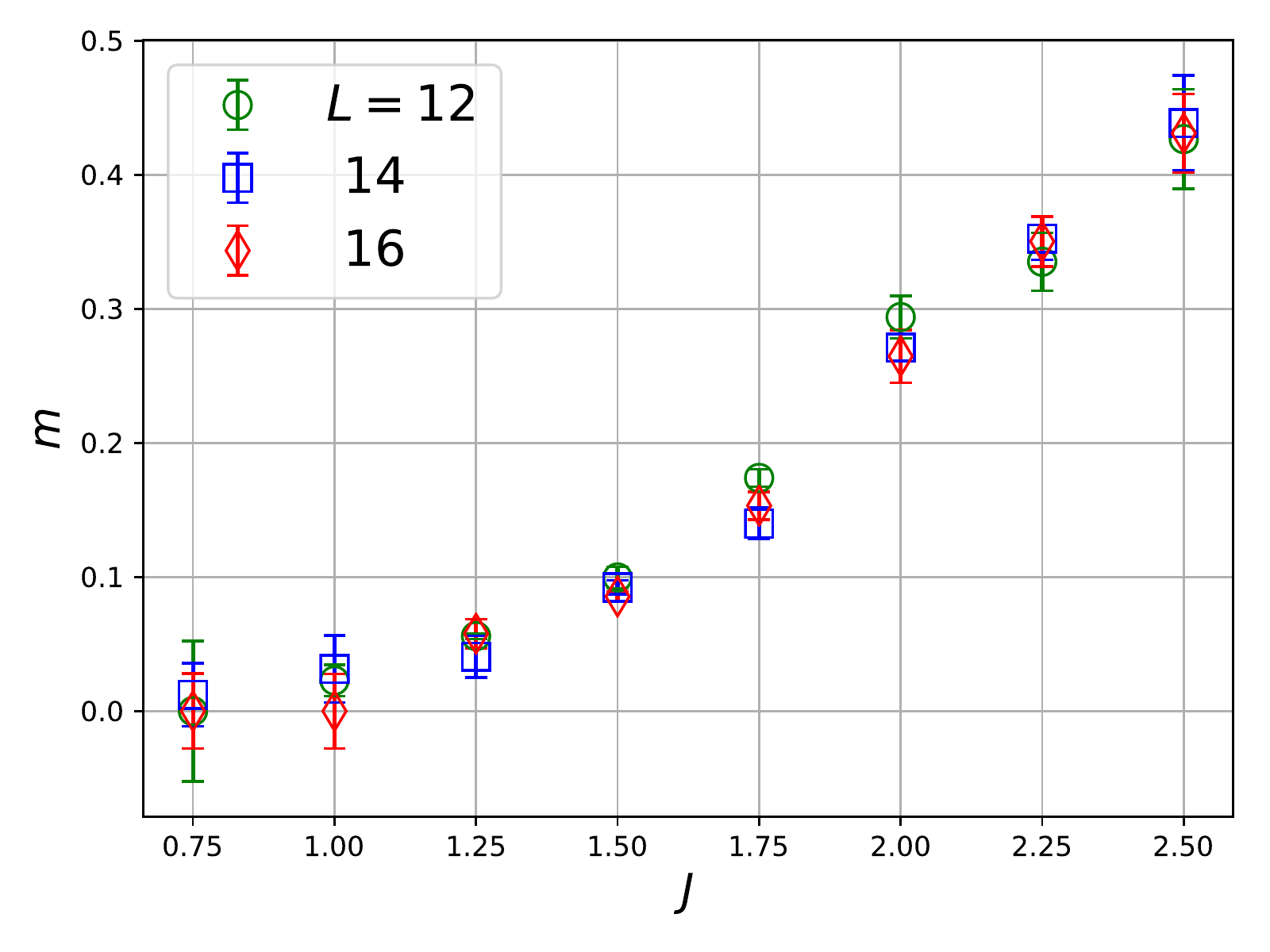}
\caption{Photon mass $m$ measured by the flux correlation
$C(\tau)=\langle \theta(\tau_0) \theta(\tau_0+\tau) \rangle\sim
\exp(-m\tau)$ with $\theta(\tau) = \sum_{\square} \sin \left(
\text{curl} \phi(\tau) \right)$. Zero photon mass is observed in
the U1D phase, finite photon mass is observed in the confined phase. Here we plot $N_f=2$ case.} \label{fig:n2fluxcorr}
\end{figure}

\begin{figure}[htp]
\includegraphics[width=\columnwidth]{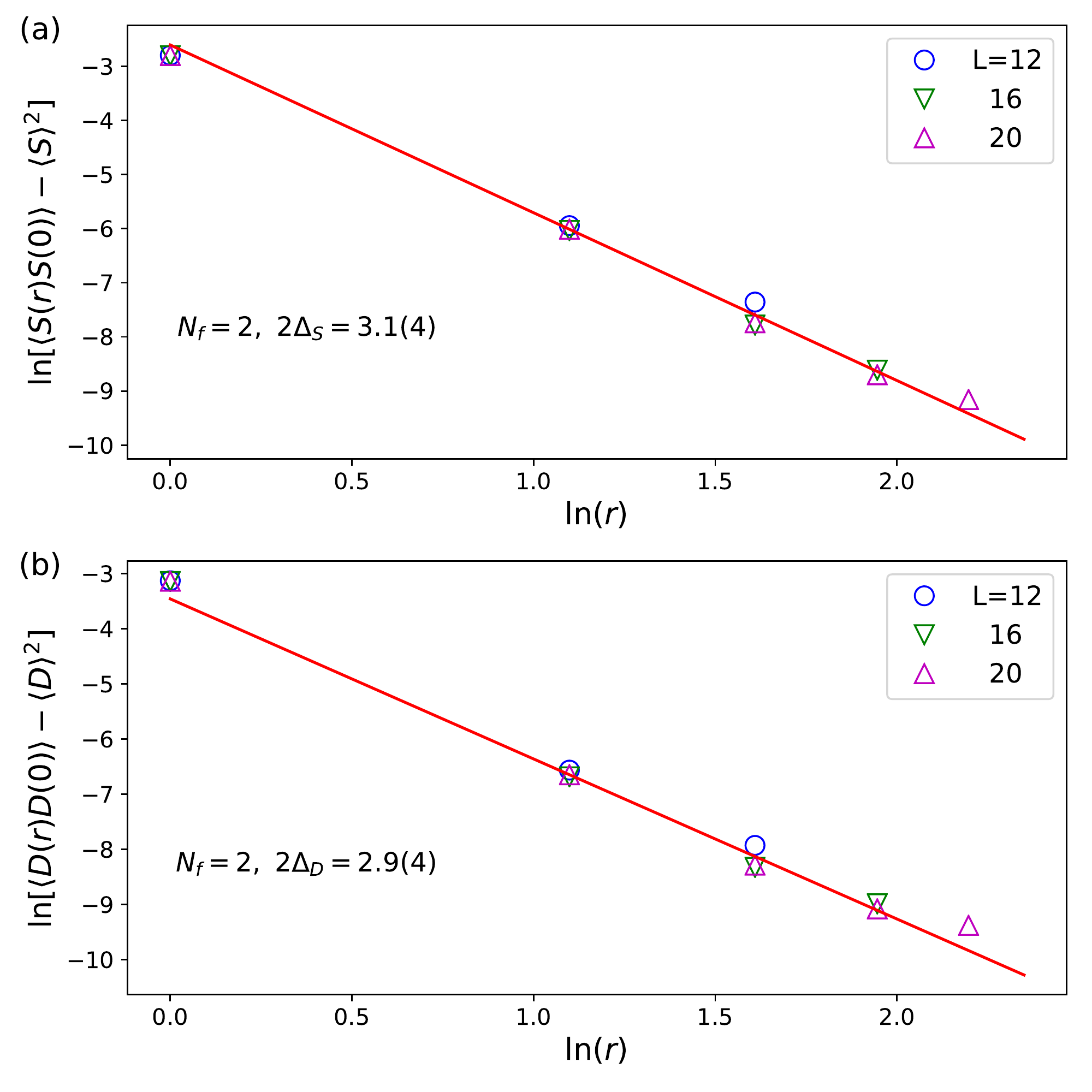}
\caption{The log-log plot of real space decay of (a) spin
correlation functions and (b) dimer correlation functions for
$N_f=2$ in the U1D phase (at $J=1.25<J_c$). The slope gives a good
estimation of the scaling dimension of spin and dimer.
Note that to avoid even-odd oscillation in the finite size data, here only the distance $r=$ odd points are plotted in the U1D phase. For other $N_f$ cases in the following, we adopt to the same strategy. } \label{fig:n2decay}
\end{figure}

A simplest way to detect the deconfinement-confinement transition
may be the Wilson loops, but it is known that in the presence of
matter field, it cannot be used to detect the topological order of
the deconfined phase (the U1D phase), such effects have been
discussed in the literature~\cite{hermele2004a,gregor2011}. One
suitable way here to demonstrate the deconfinement-confinement
transition is to show how the photon mass changes over $J$. As
soon as the matter  fields bind to form the particle-hole
condensate, we expect monopoles to proliferate and  to generate a
photon mass. The photon mass can be measured by the correlation of
flux quantity $\theta(\tau)$~\cite{burkitt1988glueballs}, which is
defined as $\theta(\tau) = \sum_{\square} \sin \left( \text{curl}
\phi(\tau) \right)$. The photon mass $m$ is related to the
correlation of $\theta(\tau)$ by $C(\tau)=\langle \theta(\tau_0)
\theta(\tau_0+\tau) \rangle \sim \exp(-m\tau)$.
Fig.~\ref{fig:n2fluxcorr} plots the estimated photon mass for
different system sizes. we find a signature of absence of photon
mass in the U1D phase and a growth of the photon
mass in the AFM phase as $J$ increases.
However, we want to point out that due to finite size effects, i.e. uncertainties in extracting the exponential decay in $\theta(\tau)$ close to the transition, the estimation of photon mass near phase transition is more qualitative than quantitative.

To further understand the properties of the deconfined phase, we
measure the real-space correlation functions in the U1D phase (at
$J=1.25<J_c$).
As shown in Fig.~\ref{fig:n2decay}(a), the spin-spin correlation shows a power-law with the power $2\Delta_S=3.1(4)$ ($\Delta_S$ is the scaling dimension of spin). Interestingly the dimer-dimer correlation function decays  with  a similar power-law with the power $2\Delta_D=2.9(4)$ ($\Delta_D$ is the scaling dimension of dimer) (Fig.~\ref{fig:n2decay}(b)).
This result sheds  light on the property of the deconfined phase, which is proposed in Refs.~\cite{Rantner2002,Hermele2005} to correspond to the algebraic spin liquid. It has the unique property that as a deconfined state emerging from  competing orders, the correlation functions of these competing orders, such as spin-spin, dimer-dimer and bond-bond, have the the same power-law decay. If the data in Fig.~\ref{fig:n2decay} were deep inside the confined phase, the decay of spin-spin and dimer-dimer correlations will be very different. For example, in the N\'eel phase, spin-spin will decay to a constant value and dimer-dimer correlation  will decay  exponentially. Therefore, our data in Fig.~\ref{fig:n2decay} provide supporting evidence of the algebraic  spin liquid behavior of the U1D in Fig.~\ref{fig:phasediagram} at $N_f=2$.

\subsubsection{$N_f=4$}
\begin{figure}[htp!]
\includegraphics[width=\columnwidth]{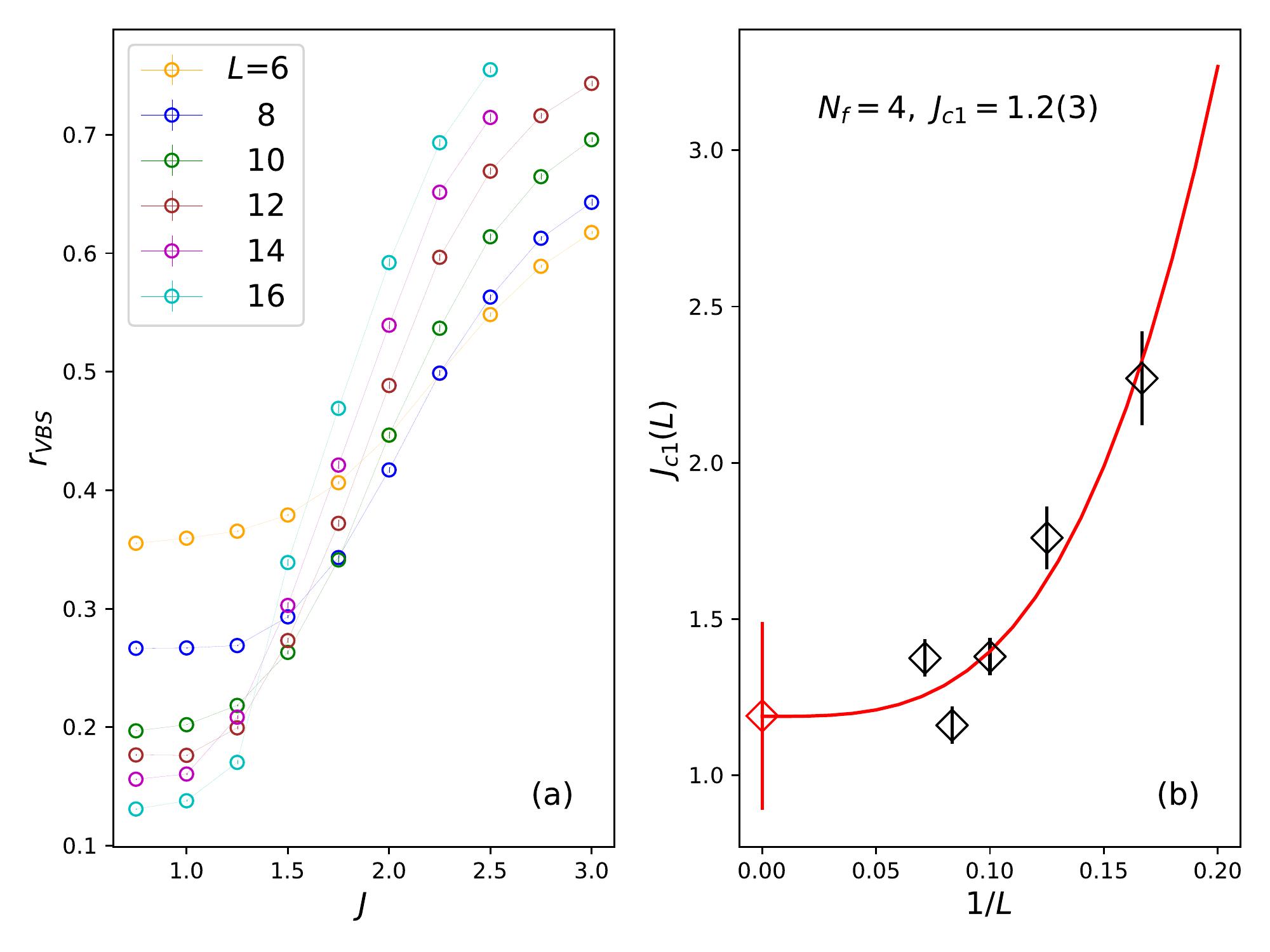}
\caption{The VBS correlation ratio through U1D to VBS transition at $N_f=4$. Here
$\beta=3L$, $\Delta \tau=0.2$. (b) The $1/L$
extrapolation of the crossings estimates the U1D to VBS transition point $J_{c1}=1.2(3)$ for $N_f=4$.}
\label{fig:n4rvbs}
\end{figure}

Next we turn to the $N_f=4$ case  where we also observe a U1D
phase at small $J$.  As shown in Fig.~\ref{fig:n4rvbs}(a), we can
follow the crossing points of the correlation ratio of the VBS
order parameter  for different system sizes,   so as to extract
(see Fig.~\ref{fig:n4rvbs}(b))  $J_{c1} = 1.2(3)$.   The data is consistent with a
continuous phase transition from the deconfined phase to the VBS
phase. Furthermore, the  flux energy per plaquette
 also supports a continuous  transition.

\begin{figure}[htp]
\includegraphics[width=\columnwidth]{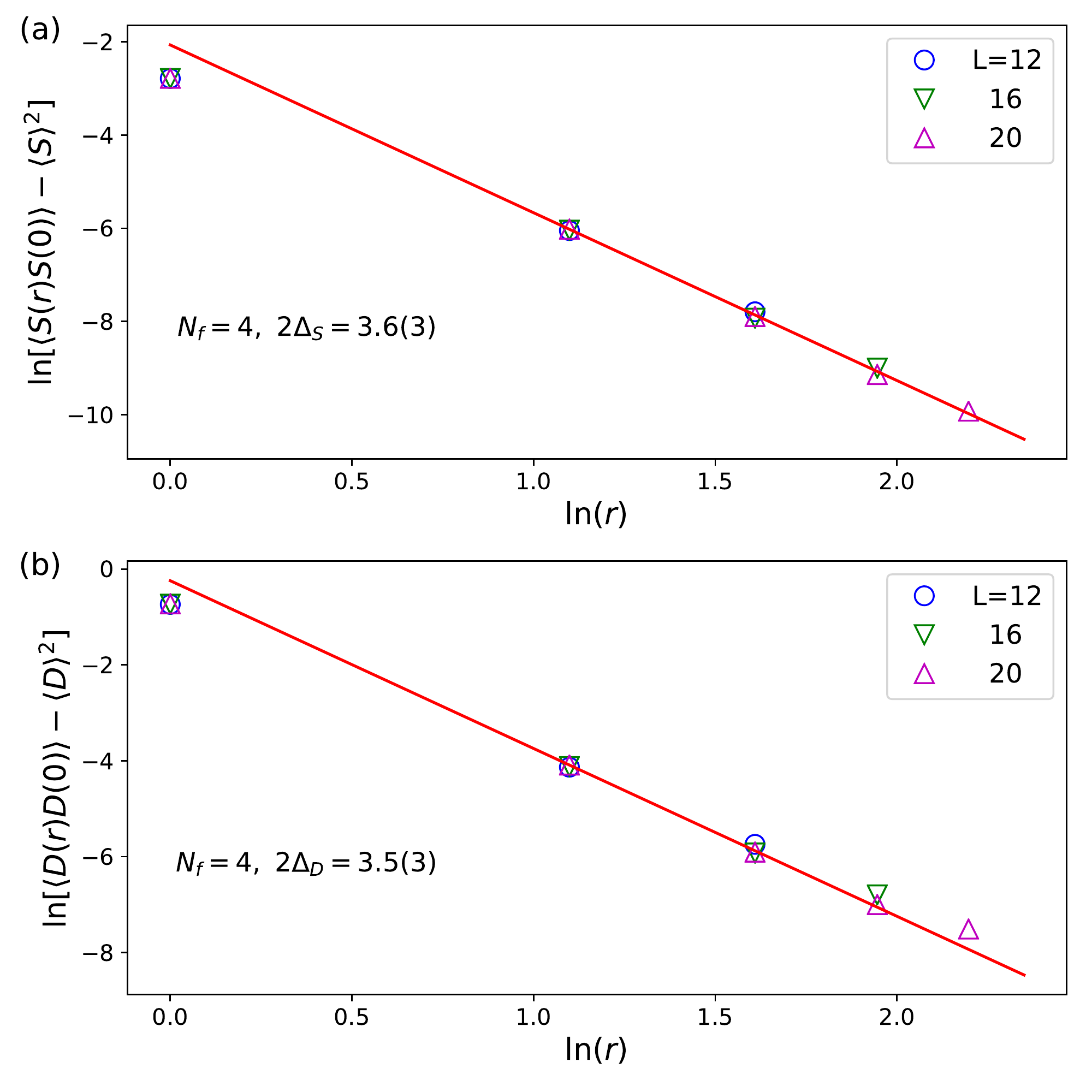}
\caption{The log-log plot of real space decay of (a) spin correlation functions and (b) dimer correlation functions for $N_f=4$ in the U1D phase (at $J=1.00<J_c$). The slope gives a good estimation of the scaling dimension of spin and dimer.} \label{fig:n4decay}
\end{figure}

Fig.~\ref{fig:n4decay} depicts the real space decay of the spin-spin and dimer-dimer correlation functions in the U1D phase for different system sizes. Again, they show similar power-law decay, and the power is estimated to be $2\Delta_S=3.6(3)$ for spin-spin correlation and $2\Delta_D=3.5(3)$ for dimer-dimer correlation.
This  power law is faster than at $N_f=2$,  and is
hence  consistent with the large-$N_f$ prediction~\cite{Rantner2002,Hermele2005,Hermele2007,Xu2007}.

Besides the power-law decay of various competing correlation functions, the situation at $N_f=4$ is even more interesting than that at $N_f=2$. As we further increase $J$,  we observe  another phase transition from  VBS to  AFM. This phase transition  is discussed  in  detail in Sec.~\ref{sec:afmvbstransition} of the main text.

\begin{figure}[t!]
\includegraphics[width=\columnwidth]{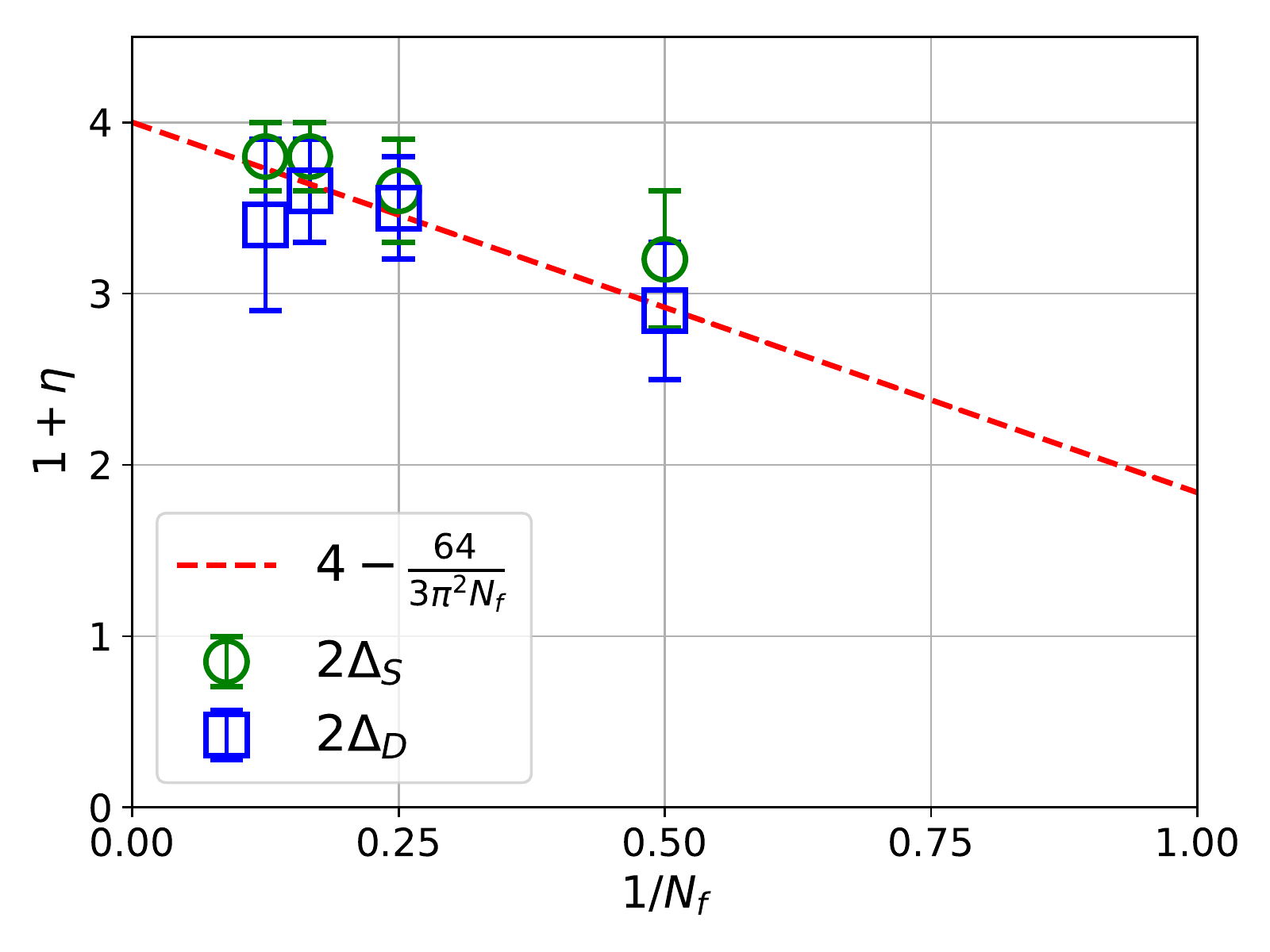}
\caption{Dimension of spin (gren circles) and dimer (blue squares) in the U1D phase  as a function of  $N_f$.  The dashed red line corresponds to the  $1/N_f$  perturbative calculation,
$1+\eta=4-\frac{64}{3\pi^2N_f}$,  taken  from Ref.~\cite{Rantner2002}. Note here  $N_f$ corresponds to the number of four-component Dirac fermions.}
\label{fig:eta}
\end{figure}
\subsubsection{Scaling dimension in U1D phase}

The U1D phase found in the small $J$ region is expected to have
the same scaling dimension for spin ($\Delta_S$) and dimer ($\Delta_D$)~\cite{Hermele2005}.
In fact, according to the
large-$N_f$ perturbative renormalization group calculation, these
correlation functions decay as
\begin{equation}
\sim r^{-(4-\frac{64}{3\pi^2 N_f})}
\end{equation}
with correction at
$O(\frac{1}{N^2_f})$~\cite{Rantner2002,Hermele2005,Hermele2007,Xu2007}. Notice that in our case, $N_f$ is the number of fermion flavors on the
lattice, while in
Refs.~\cite{Rantner2002,Hermele2005,Hermele2007,Xu2007}, $N_f$ is
the number of two-component Dirac fermions, which is twice of our
$N_f$ due to momentum valley degeneracy.

We now compare this theoretical expectation to our numerical
simulation results.
Fig.~\ref{fig:eta} presents a summary plot of the power law we obtain at $N_f=2$, $4$, $6$, and $8$. The results for $N_f=6$ and $N_f=8$ are detailed in Appendix~\ref{sec:nf6and8}. It is remarkable to see that our data perfectly matches the aforementioned $1/N_f$ perturbative expression.

\subsubsection{Theory for confinement transition}

We find a U1D to AFM phase transition for the  $N_f=2$ case and  a
U1D to VBS phase transition for the $N_f=4$, $6$ and $8$ cases.
These phase transitions should belong to the
QED$_3$-Gross-Neveu-O(3) or XY transitions~\cite{Janssen2017},
depending upon the order parameters in the confined phases. For
example, at least with large enough (but still
finite) $N_f$ when the higher order fermion interactions are
clearly irrelevant, the transition between U1D to VBS phase can
be described by the following action:
\begin{eqnarray}
\mathcal{S} = \int d^2x d\tau \sum_{j = 1}^{2N_f} \bar{\psi}_j
\gamma \cdot (\partial - i a)\psi_j + u \vec{\phi} \cdot
\bar{\psi} \vec{\mu} \psi \nonumber\\
+ |\partial \vec{\phi}|^2 + r
|\vec{\phi}|^2 + g|\vec{\phi}|^4,
\end{eqnarray}
where $\psi=(\psi_1, \cdots, \psi_{2N_f})^T$ has $2N_f$
components, $\vec{\phi}$ is a $O(2)$ vector in which the VBS order
parameter is embedded. $\vec{\mu} = (\mu^x, \mu^y)$ are two $2N_f
\times 2N_f$ matrices in the fermion flavor space.
$\bar{\psi}\mu^x\psi$, $\bar{\psi}\mu^y\psi$ are two fermion mass
operators that correspond to the VBS in $x$ and $y$ directions
respectively. When $r > 0$, $\vec{\phi}$ is gapped out, and the
system is in the U1D phase due to the screening of massless
fermions to the gauge field. When $r < 0$, $\vec{\phi}$ condenses,
the fermions are gapped out, then the compact gauge field is in
the confined phase. In the U1D phase, the VBS and
antiferromagnetic order parameters should have the same scaling
dimension, due to the enlarged SU($2N_f$) symmetry in the low
energy field theory, consistent with our data in
Fig.~\ref{fig:n8decay} and
Figs.~\ref{fig:n2decay},\ref{fig:n4decay} and \ref{fig:n6decay} in
Appendix~\ref{sec:phaseboundaries}; while at the critical point $r
= 0$, these two order parameters still have power-law correlation
functions, but with different scaling dimensions, due to the loss
of the SU($2N_f$) symmetry in the infrared.

\subsubsection{AFM-VBS transition at $N_f = 4$}
\label{sec:afmvbstransition}

\begin{figure}[h!]
\includegraphics[width=\columnwidth]{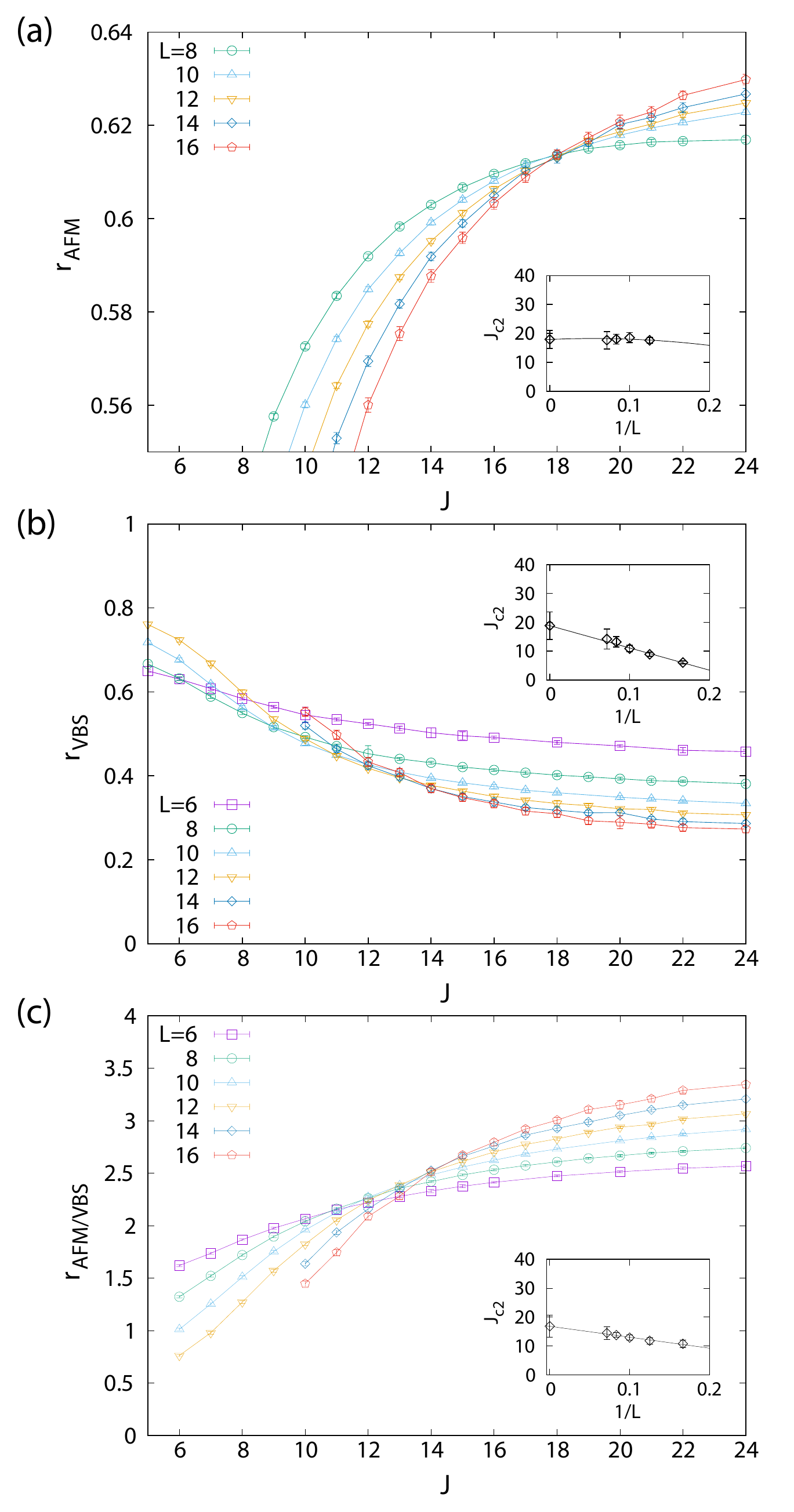}
\caption{(a) Antiferromagnetic correlation ratio $r_{\text{AFM}}$
for $N_f=4$. Here $\beta=3L$, $\Delta \tau=0.2$. Insets shows the
$1/L$ extrapolation of the crossing point in $r_{\text{AFM}}$ and
$J_{c2}=18(3)$. (b) VBS correlation ratio $r_{\text{VBS}}$ for
$N_f=4$. The inset is the $1/L$ extrapolation of the crossing
point in $r_{\text{VBS}}$ and $J_{c2}=19(5)$. This is consistent with $J_{c2}$
obtained from $r_{\text{AFM}}$ in (a). (c) AFM/VBS correlation
ratio $r_{\text{AFM/VBS}}$ for $N_f=4$. The inset is the $1/L$
extrapolation of the crossing point in $r_{\text{AFM/VBS}}$ and
$J_{c2}=17(4)$.
}
\label{fig:n4rafm}
\end{figure}

The situation  at $N_f=4$ is even more interesting. As we further
increase $J$,   we observe  another quantum phase transition from
VBS to AFM phase. This transition is consistently revealed in
three steps in Fig.~\ref{fig:n4rafm}.

Fig.~\ref{fig:n4rafm} (a) shows the $r_{\text{AFM}}$ correlation
ratio and clearly there is a crossing point signifying the
establishment of the AFM long range order. The inset shows the
$1/L$ extrapolation of the crossing point and gives rise to
$J_{c2}=18(3)$. Fig.~\ref{fig:n4rafm} (b) is the $r_{\text{VBS}}$
correlation ratio and the crossing point in it signifies the
vanishing of the VBS order. Inset of Fig.~\ref{fig:n4rafm} (b)
gives rise to $J_{c2}=19(5)$, consistent with the onset of the AFM
order in Fig.~\ref{fig:n4rafm} (a).

The transition from VBS to AFM deserves more attention, apparently
the data in Fig.~\ref{fig:n4rafm} suggest a continuous transition,
and if it were the case, there is then the possibility that the
critical point will acquire larger symmetry group than that in the
model in Eq.~\eqref{eq:hamiltonian}. In the case of $Z_2$ gauge
field coupled to fermion, as shown in
Refs.~\cite{Assaad2016,Gazit2018}, two similar situations with
emergent continuous symmetry are also investigated. In the first
case, it is at $N_f=3$ that a continuous VBS to AFM phase
transition occurs~\cite{Assaad2016}, and in the second case, it is
the deconfine-confine phase transition itself at
$N_f=2$~\cite{Gazit2018}. Our phase transition at $J_{c2}$ is
closer to the former. To further understand the nature of
transition from VBS to AFM, we plot the ratio of AFM structure
factor and VBS structure factor,
\begin{equation}
r_{\text{AFM/VBS}}= \frac{\chi_S(\vec{X})}{\chi_D(\vec{M})}.
\end{equation}

The results are shown in Fig.~\ref{fig:n4rafm} (c), indeed there
is a crossing point in the $r_{\text{AFM/VBS}}$, and the $1/L$
extrapolation in the inset of Fig.~\ref{fig:n4rafm} (c) gives rise
to $J_{c2}=17(4)$, very consistent with the $J_{c2}$ obtained from
the crossings of $r_{\text{AFM}}$ in Fig.~\ref{fig:n4rafm} (a) and
$r_{\text{VBS}}$ in Fig.~\ref{fig:n4rafm} (b). This transition is
similar to the AFM-to-VBS transition in the $SU(4)$ J-Q
model~\cite{Lou2009}, where a continous transition is observed.
The transition in that case can be described by a NCCP$^{N-1}$
description with $N=4$~\cite{Kaul2012}, and it is shown that the
monopoles are irrelevant at this fix point~\cite{Block2013} and
therefore a deconfined quantum critical point~\cite{Senthil2004}
is realized. However, in Ref.~\cite{Kaul2012}, on sublattice A of
the lattice there is a fundamental representation of SU(4), while
on sublattice B there is a anti-fundamental representation, which
is different from our case.

In our case, with $N_f = 4$ there is effectively a self-conjugate
representation of the SU(4) group on every site, thus the field
theory for the AFM-to-VBS transition is different from
Ref.~\cite{Kaul2012}. According to Ref.~\cite{sachdevread}, the
AFM N\'{e}el order in this case has the following grassmanian
ground state manifold $\mathcal{M}$:
\begin{equation}
\mathcal{M} = \frac{U(4)}{U(2)\times U(2)}.
\end{equation}
To describe this antiferromagnetic state, one can either introduce
$N_f = 4$ flavors of fermionic spinons with half-filling, or
introduce two color species of bosonic spinons $z_{\alpha,a}$
($\alpha = 1 ,\cdots 4$, $a = 1,2$), and couple them to a U(2)
gauge field (to describe the simplest N\'{e}el order of SU(2)
spins, we only need one two component of bosonic spinon coupled
with a U(2) gauge field, as Ref.~\cite{Senthil2004}). The U(2)
gauge constraint will guarantee that on every site there are fixed
number of spinons, and the color space is fully antisymmetric,
thus on every site the SU(4) spin is automatically in an
antisymmetric self-conjugate representation. Then the field theory
for the N\'{e}el-VBS transition is
\begin{equation}
\mathcal{S} = \int d^2x d\tau \ |(\partial - i a - i \sum_{l =
1\cdots 3} a^l \tau^l)z|^2 + r |z|^2 + \cdots
\end{equation}
where $a_\mu$ and $a^l_{\mu}$ are gauge fields corresponding to
the U(1) and SU(2) subgroups of U(2). Note that these gauge fields
are ``emergent" gauge fields, which are different from the
explicit gauge field in our original simulation.

When $r < 0$, $z_{\alpha,a}$ condenses, and leads to the
antiferromagnetic state with ground state manifold
$\frac{U(4)}{U(2)\times U(2)}$. When $r > 0$, $z_{\alpha,a}$ is
gapped out, and the gauge fields will be confined. Here we assume
that the U(1) compact gauge field $a_\mu$ still has the
quadru-monopole proliferation, which leads to the VBS phase like
the original deconfined QCP theory for the SU(2)
spins~\cite{Senthil2004}.

One of the crucial properties of the deconfined QCP is the
``intertwinement" between order parameters on two sides of the
transition, which can be captured by a topological term which
treats the N\'{e}el and VBS order parameters on an equal
footing~\cite{senthilfisher}. In the current case with SU(4) spin
symmetry, one can also introduce a topological term that captures
the ``intertwinement" between the the SU(4) N\'{e}el and VBS
orders with a topological term. To do this, we need to embed both
the N\'{e}el and VBS order parameters into a larger manifold. One
way to parameterize the N\'{e}el order manifold is $\mathcal{N} =
U^\dagger \Omega U$, where $\Omega$ is a $4\times 4$ diagonal
matrix $\Omega = \mathrm{diag}(\mathbf{1}_{2\times 2}, -
\mathbf{1}_{2\times 2})$, and $U$ is a SU(4) matrix. $\mathcal{N}$
is a $4\times 4$ Hermitian matrix with constraint $ \mathcal{N}^2
= 1$.

Now we introduce a larger $8\times 8$ matrix $\mathcal{P}$ which
includes both the N\'{e}el and VBS order parameters:
\begin{equation}
\mathcal{P} = \cos(\theta) \mathcal{N} \otimes \tau^z +
\sin(\theta) \mathbf{1}_{4\times 4} \otimes \left( V_x \tau^x +
V_y \tau^y \right),
\end{equation}
where $(V_x, V_y)$ is a two component order parameter for the VBS
phase, and $V_x^2 + V_y^2 = 1$. The order parameter $\mathcal{P}$
unifies the SU(4) N\'{e}el and VBS order parameters, just like the
O(5) vector order parameter introduced in
Ref.~\cite{senthilfisher}.

The topological term that captures the ``intertwinement" between
N\'{e}el and VBS order parameters is a Wess-Zumino-Witten term:
\begin{equation}
\mathcal{S}_{wzw} \sim \int d^2x d\tau \int_0^1 du \
\epsilon_{\mu\nu\rho\sigma}\mathrm{tr}[\mathcal{P} \partial_\mu
\mathcal{P}
\partial_\nu \mathcal{P} \partial_\rho \mathcal{P} \partial_\sigma
\mathcal{P}]. \label{wzw}
\end{equation} Using the same technique in
Ref.~\cite{groversenthil}, one can show that at the vortex core of
the VBS order parameter, there is a spinon with self-conjugate
representation, which is consistent with intuition. In fact, the
O(5) WZW term introduce in Ref.~\cite{senthilfisher} can be
written in the same form as Eq.~\eqref{wzw}, as long as we replace
$\mathcal{P}$ in Eq.~\eqref{wzw} by a $4\times 4$ Hermitian matrix
order parameter $\mathcal{P} = \vec{n} \cdot \vec{\Gamma}$, where
$\vec{\Gamma}$ are five Gamma matrices, and $\vec{n}$ is the O(5)
vector introduced in Ref.~\cite{senthilfisher}.

This topological term can be viewed as the low-energy effective
field theory of the $\pi-$flux state of the SU(4) antiferromagnet,
which again is described by a QED$_3$ with eight flavors of Dirac
fermions~\cite{Hermele2005}, but again the gauge field of this
QED$_3$ is an emergent gauge field which is different from the
gauge field introduced in the original model that we simulate. The
WZW term Eq.~\eqref{wzw} can be derived by coupling the $8\times
8$ matrix order parameter $\mathcal{P}$ to the eight flavors of
Dirac fermions of the $\pi-$flux state, and integrate out the
fermions following the standard procedure of Ref.~\cite{abanov}.

Our data, the crossing of $r_{\text{AFM}/\text{VBS}}$ in
Fig.~\ref{fig:n4rafm} (c), suggest that the AFM and VBS order
parameters have the same scaling dimension at this transition,
which is consistent with the emergent SU(8) symmetry of the
$\pi-$flux state of the SU(4) antiferromagnet. The large SU(8)
symmetry, if indeed exists at the AFM-VBS transition, will ensure
that many other order parameters have the same scaling dimension
as AFM and VBS order parameters~\cite{Hermele2005}. These order
parameters will also have similar fractionalization dynamical
signatures in their spectral functions as AFM and VBS order
parameters.

\section{Conclusions}
\label{sec:discuss}

Using large scale DQMC,  we investigated the compact $U(1)$ gauge
field theory coupled to Dirac fermion matter fields in $(2+1)$D
and variable flavor number $N_f$:  i.e.  cQED$_3$. With our
simulations we mapped out the entire ground state phase diagram in
the flavor $N_f$ and gauge field fluctuation $J$ strength plane.
Our results are summarized in Fig.~\ref{fig:phasediagram}(a).

Most importantly, signatures supporting stable
U(1) deconfined (U1D) phases were discovered at $N_f=8$ and $6$,
and evidence of the U1D phase at $N_f=4$ and $2$ were also found.
The properties of the deconfined phase are consistent with the
proposal of algebraic spin liquid, in which, various competing
orders (AFM order and VBS order, for example) all have algebraic
correlation  with identical power-laws in
real-space. The decay power is found to quantitatively converge to
the large-$N_f$
predictions~\cite{Rantner2002,Hermele2005,Hermele2007,Xu2007}.

The transition between the deconfined and confined phases at
various  $N_f$ were determined using  the RG invariant correlation
ratios. At $N_f=2$ the transition occurs between the U1D  and AFM
phases. Since the AFM    corresponds to O(3) symmetry breaking,
the  critical theory should be described by  the
QED$_3$-Gross-Neveu O(3)  universality class.  In contrast,  at
larger values of $N_f$  the ordered phases  corresponds to VBS.
The dynamical generation of the  two VBS mass terms is described
by the QED$_3$-Gross-Neveu O(2) universality class. As far as we
know, the QED$_3$-Gross-Neveu O(2) or O(3) transition have not
been investigated numerically before in an unbiased simulation. It
is certainly worthwhile to further carefully study the critical
properties of these transitions via QMC simulations and compare
with  future analytical calculations. In particular,  the
QED$_3$-Gross-Neveu Ising transition has been investigated
recently with perturbative RG
calculations~\cite{Janssen2017,Ihrig2018}.

Aside from the QED$_3$-Gross-Neveu transitions, we have found
evidence for a direct and continuous transition between the AFM
and VBS states in the confined region of the phase diagram at $N_f
= 4$. Since we have on average two fermions per site, we should
consider the antisymmetric self-conjugate representation of the
SU(4) group. We have presented various theoretical descriptions of
this putative deconfined quantum phase transition in terms of
multi-flavored spinons coupled to emergent U(1) and SU(2) gauge
fields. We also discussed the effective low energy field theory
with a topological term that captures the intertwinement between
the magnetic and VBS orders, and its connection to the $\pi-$flux
state of the SU(4) spin system discussed in
Ref.~\cite{Hermele2005}. Numerical support for emergent symmetries
was provided. In the future, measurements of the conserved current operators related with such emergent continuous symmetries~\cite{Ma2018Current} can be performed.

Finally, the confining phase transitions in our model, as well as
the possible deconfined quantum critical point at $N_f=4$, will
have distinct and very interesting dynamical properties in their
spectral functions, that can be further explored in QMC
simulations.   Such calculations   provide experimentally
accessible signatures of exotic states of matter where emergent
gauge fields, fractionalized excitations, can be traced. Similar
attempts have recently been applied to the deconfined quantum
critical point in pure spin model~\cite{Ma2018}, emergent $Z_2$
spin liquid at (2+1)D~\cite{GYSun2018} and $U(1)$ spin liquid at
(3+1)D~\cite{Huang2018} and the $Z_2$ counterpart of our
model~\cite{Assaad2016}. In the present cQED$_3$ model, dynamical
measurements in the QMC simulation plus state-of-art analytical
continuation~\cite{Sandvik2015,Shao2017b,GYSun2018} can help to
reveal more fundamental physical understanding of these exotic
quantum phase transitions.

\begin{acknowledgements}
The authors thank Lukas Janssen, Ribhu Kaul, Steven Kivelson, Thomas Lang, Srinivas Raghu, Subir Sachdev, Michael Scherer, Yi-Zhuang You, Ashvin Vishwanath, Chong Wang for helpful discussions. FFA
would like to thank T. Grover and Zhenjiu Wang for discussions on
related projects. XYX, YQ, LZ, CX acknowledge the hospitality of
the International Collaboration Center at Institute of Physics,
Chinese Academy of Sciences, this work is finalized during the
program "International Workshop on New Paradigms in Quantum
Matter" at the Center. XYX is thankful for the support of HKRGC through through C6026-16W, 16324216 and 16307117.
YQ acknowledges support from Minstry of Science and Technology of China under grant numbers 2015CB921700, and from National Science Foundation of China under grant number 11874115.
ZYM acknowledges supports from the National Key R\&D Program (2016YFA0300502), the Strategic Priority Research Program of
CAS (XDB28000000), and the National
Science Foundation of China (11574359, 11674370). FFA thanks the
DFG research unit FOR1807 for financial support. LZ is supported
by the National Key R\&D Program of China (Grant No.
2018YFA0305802) and the Key Research Program of Chinese Academy of
Sciences (Nos. XDPB08-4 and XDB28040200). We thank the Center for
Quantum Simulation Sciences in the Institute of Physics, Chinese
Academy of Sciences, the Tianhe-1A platform at the National
Supercomputer Center in Tianjin and Tianhe-2 platform at the
National Supercomputer Center in Guangzhou for their technical
support and generous allocation of CPU time.
\end{acknowledgements}

\appendix
\section{Connection to high energy lattice cQED$_3$ action}
As we discussed in the Sec.~\ref{sec:model}, after the path
integral of the rotor degrees of freedom in a rotor model with
fermion in Eq.~\eqref{eq:hamiltonian}, the
action of cQED$_3$ coupled to fermionic matter is obtained explicitly. In the high energy
lattice cQED$_3$ action, the Lagrangian for pure gauge field part
takes the form
\begin{equation}
L_\phi = K_{\tau} \sum_{\langle ij \rangle} \left(
\cos(\phi_{ij}(\tau+1)-\phi_{ij}(\tau)) \right) + K_{r}
\sum_{\square}\cos \left( \text{curl} \phi \right)
\label{eq:hep-lagrangian}
\end{equation}
where $K_{\tau}<0$ and $K_r<0$ with $|K_{\tau}|=|K_r|$. Compared with the
Lagrangian defined in Eq.~\eqref{eq:lagrangian}, we study the case
$K_{\tau}<0$ and $K_r>0$ with $|K_{\tau}|\ne |K_r|$. As we can always
rescale space and time to restore the Lorentz symmetry, the
difference between $|K_{\tau}| = |K_r|$ and $|K_{\tau} \ne |K_r|$ is
trivial. Actually, our model can be exactly mapped to the case of
$K_r < 0$ and the fermion hopping with a staggered phase factor as follows:
\[\phi_{i,i+\hat x}\rightarrow\phi_{i,i+\hat x}+m_y(i)\pi,\]
where $m_y(i)$ is 1 (0) if the $y$-coordinate of $i$ is odd
(even), respectively. Therefore, our convention is equivalent to
the high energy lattice cQED$_3$ action.

As we mentioned in the main text, the Dirac fermion in our model
is realized, because the $K_r$ term prefers $\pi$-flux through
each plaquette, and the $\pi$-flux doubles the unit cell.
Following the standard literature such as Ref.~\cite{Hermele2005},
if we start with $N_f$ flavors of one-component fermions on the
lattice (like Eq.~\eqref{eq:hamiltonian}), at low energy there
will be $2N_f$ flavors of 2-component Dirac fermions.

\section{Pseudo-unitary group $SU(n,m)$ and the absence of sign problem}
\label{sec:psu}
As we discussed in the main text, the fermion determinant for one flavor is $\det \big( I_{n+m}+\prod_{z=1}^{L_\tau}B_z \big)$, where $n$ and $m$ are the numbers of sites in the two sublattices, and $I_{n+m}$ denotes the $(n+m)$-dimensional identity matrix. $B_{z}=e^{h_{z}}$, where $h_{z}$ has the following structure,
\begin{equation}
h_{z}=
\begin{pmatrix}
0_{n} & T_{z} \\
T_{z}^{\dag} & 0_{m}
\end{pmatrix},
\end{equation}
and $T_{z}$ is the hopping matrix between different sublattices.

$B_{z}$ matrices satisfy (1) $B_{z}^{\dag}\eta B_{z}=\eta$, where
$\eta=\mathrm{diag}(I_{n},-I_{m})$, and (2) $\mathrm{det}B_{z}=1$,
thus their products generate the pseudo-unitary group
$\mathrm{SU}(n,m)$.

\emph{Theorem.} For any $D\in \mathrm{SU}(n,m)$, $\det
(I_{n+m}+D)\in \mathbb{R}$.

\emph{Proof.} First, suppose that $\lambda$ is an eigenvalue of $D$, $Dv=\lambda v$, then $D^{\dag}\eta v=\lambda^{-1}\eta v$, and $D^{T}\eta v^{*}=(\lambda^{*})^{-1}\eta v^{*}$, hence $(\lambda^{*})^{-1}$ is an eigenvalue of $D^{T}$, and is thus also an eigenvalue of $D$.

Denote the eigenvalues of $D$ by $\lambda_{i}$, $1\leq i\leq n+m$, then $\det (I_{n+m}+D)=\prod_{i}(1+\lambda_{i})$. We then treat the eigenvalues on the unit circle and those not on the unit circle separately. For those not on the unit circle,
\begin{equation}
\prod_{i,|\lambda_{i}|\ne1}(1+\lambda_{i})=\prod_{i,|\lambda_{i}|<1}(1+\lambda_{i})(1+(\lambda_{i}^{*})^{-1})=\prod_{i,|\lambda_{i}|<1}\frac{|1+\lambda_{i}|^{2}}{\lambda_{i}^{*}}.
\end{equation}
For those on the unit circle, denoting $\lambda_i = e^{i\theta_i}$, $-\pi<\theta_i\leq \pi$, we have
\begin{equation}
\prod_{i,|\lambda_{i}|=1}(1+\lambda_{i})=\prod_{i,|\lambda_{i}|=1}2\cos(\theta_{i}/2)e^{i\theta_{i}/2}.
\end{equation}
Therefore, we find
\begin{equation}
\begin{split}
(\det(I_{n+m}+D))^{2}&=|\det(I_{n+m}+D)|^{2}\prod_{i}\frac{\lambda_{i}}{|\lambda_{i}|}\\
&=|\det(I_{n+m}+D)|^{2}> 0,
\end{split}
\end{equation}
hence $\det(I_{n+m}+D)\in \mathbb{R}$.

This theorem implies that for an even number of flavors fermions, the
model is free of sign problem for any hopping matrices $T_{z}$.
For instance, this is true for both abelian and nonabelian gauge
fields.

It is easy to find an example
\begin{equation}
D=
\begin{pmatrix}
-\sqrt{2}& 1\\
1& -\sqrt{2}
\end{pmatrix}
\in\mathrm{SU}(n,m)
\end{equation}
such that $\det (I+D)<0$. Therefore, the absence of sign problem
does not hold for an odd number of fermions in general. However,
it does hold for models with fermions coupled to $\mathbb{Z}_{2}$
gauge fields \cite{Assaad2016, Gazit2017, Gazit2018}.

\section{Confinement transition for $N_f=6$ and $N_f=8$}
\label{sec:nf6and8}
In this section we discuss the results for $N_f=6$ and $N_f=8$.

\begin{figure}[htp]
\includegraphics[width=\columnwidth]{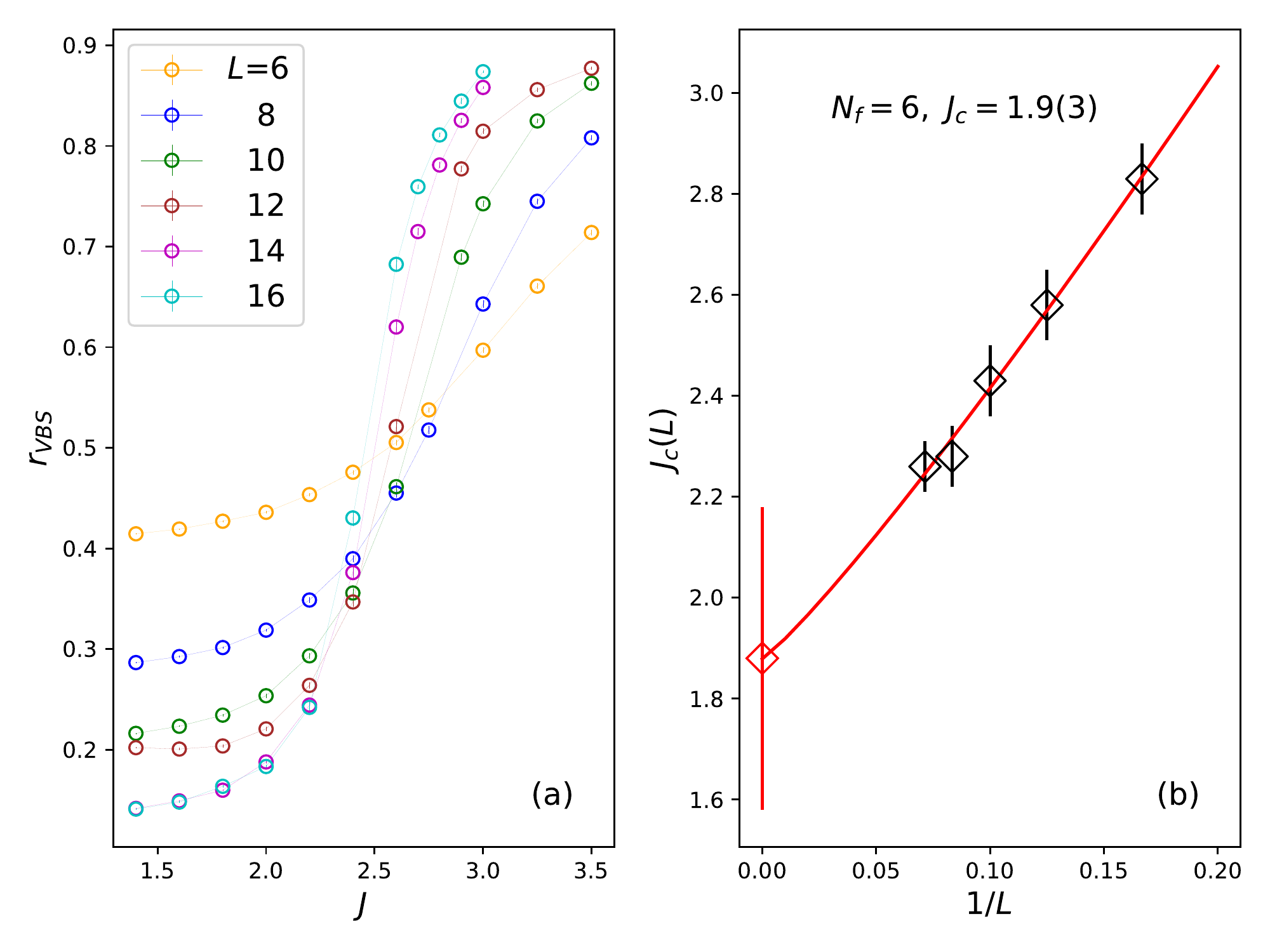}
\caption{The VBS correlation ratio through U1D to VBS transition at $N_f=6$. Here $\beta=2L$, $\Delta \tau=0.1$. (b) The $1/L$ extrapolation of the crossings estimates the U1D to VBS transition point at $J_c=1.9(3)$ for $N_f=6$.}
\label{fig:n6rvbs}
\end{figure}

\begin{figure}[htp]
\includegraphics[width=\columnwidth]{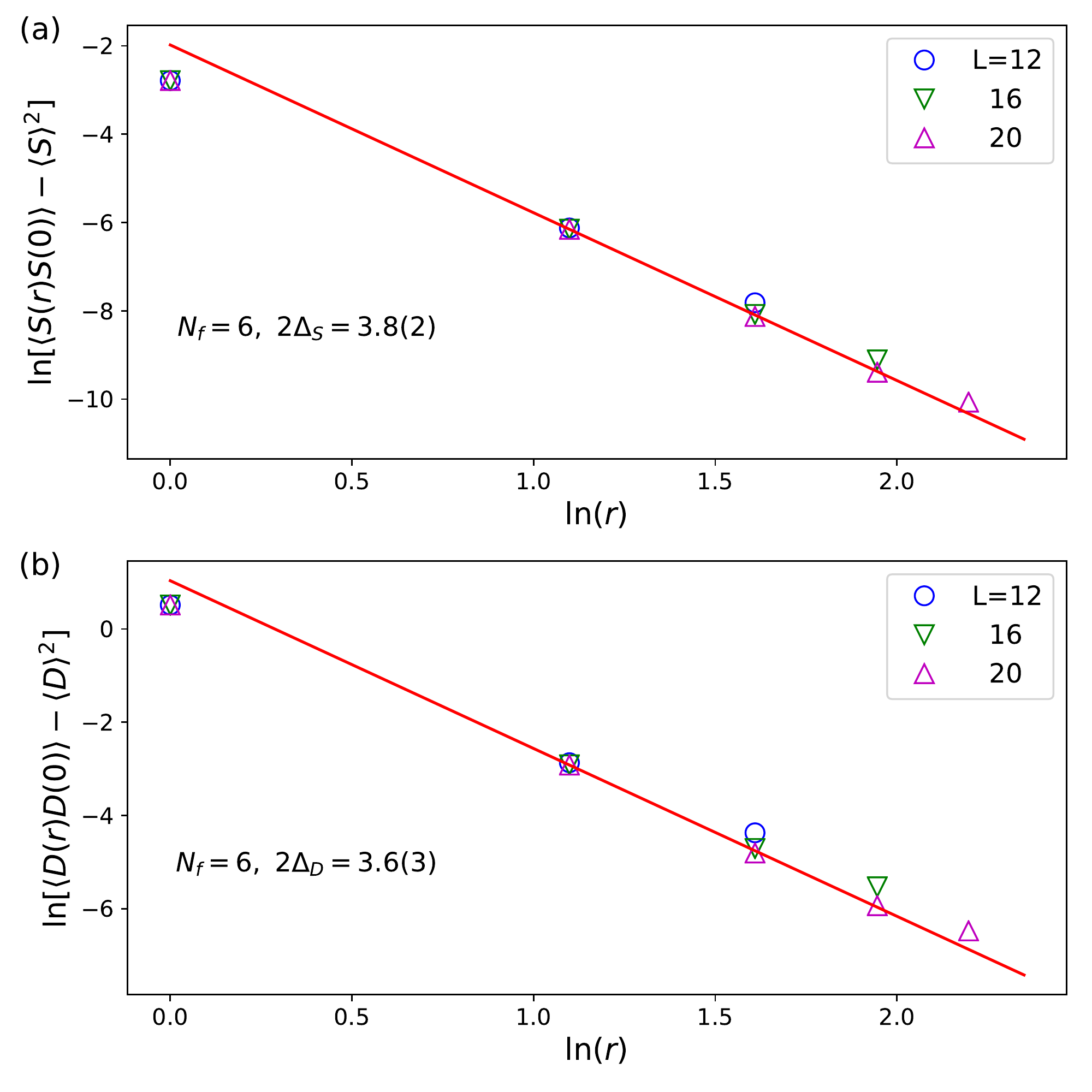}
\caption{The log-log plot of real space decay of (a) spin correlation functions and (b) dimer correlation functions for $N_f=6$ in the U1D phase (at $J=1.40<J_c$). The slope gives a good estimation of the scaling dimension of spin and dimer.}
\label{fig:n6decay}
\end{figure}

\begin{figure}[h!]
\includegraphics[width=\columnwidth]{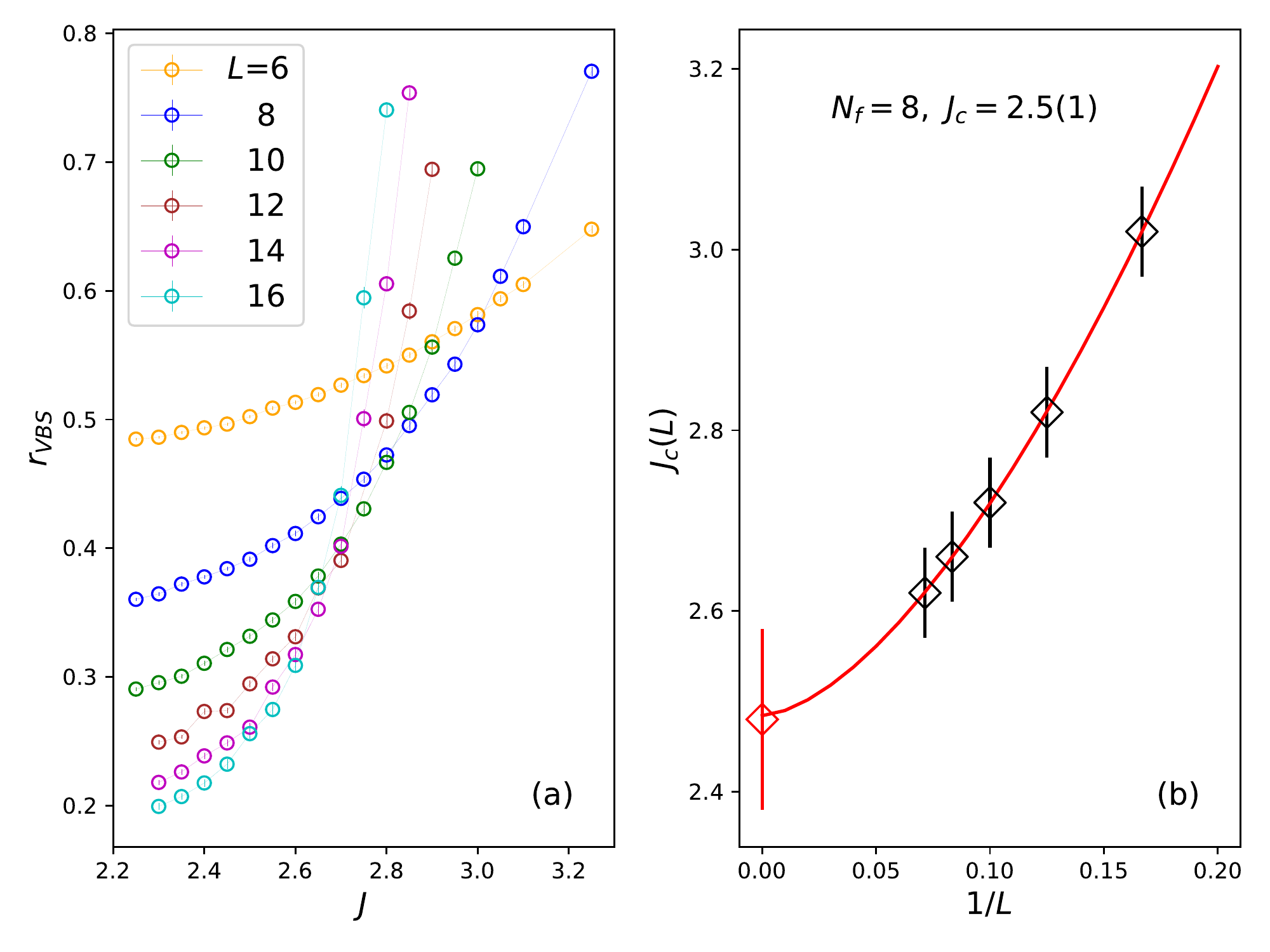}
\caption{The VBS correlation ratio through U1D to VBS transition at $N_f=8$. Here $\beta=2L$,  $\Delta \tau=0.1$. (b) The $1/L$ extrapolation of the crossings estimates U1D to VBS transition point at $J_c=2.5(1)$ for $N_f=8$.}
\label{fig:n8rvbs}
\end{figure}

As shown in Fig.~\ref{fig:n6rvbs} and Fig.~\ref{fig:n8rvbs}, and the corresponding insets,  we estimate   $J_c=1.9(3)$ for $N_f=6$ and $J_c=2.5(1)$ for $N_f=8$.  Again, the data are consistent with a  continuous transition between the deconfined  UID and confined VBS phases. Correspondingly, the flux energy per plaquette
behaves as smooth function across the critical point.

\begin{figure}[h!]
\includegraphics[width=\columnwidth]{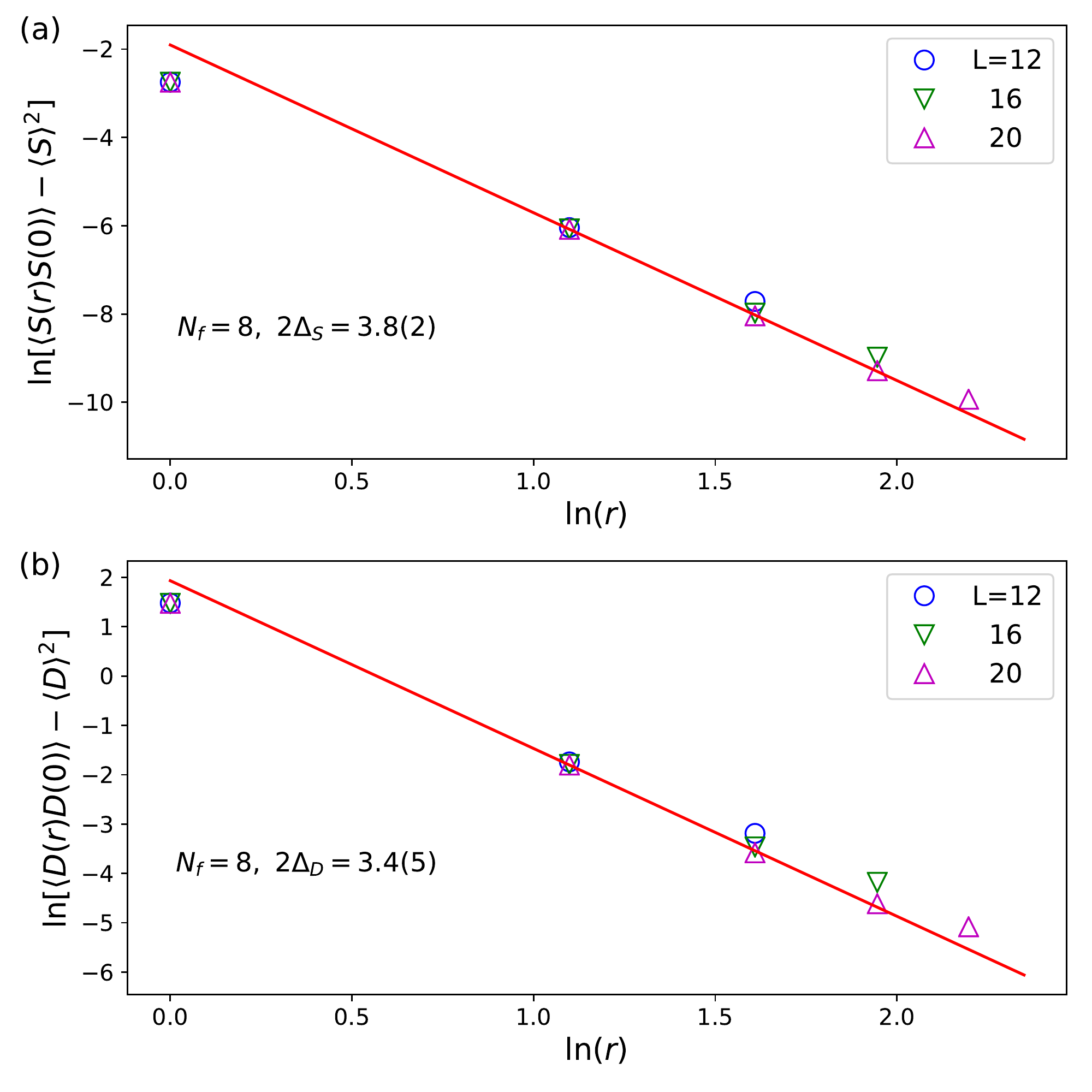}
\caption{The log-log plot of real space decay of (a) spin correlation functions and (b) dimer correlation functions for $N_f=8$ in the U1D phase (at $J=2.00<J_c$). The slope gives a good estimation of the scaling dimension of spin and dimer.}
\label{fig:n8decay}
\end{figure}

More interestingly, we plot the spin-spin and dimer-dimer correlation function in real-space for $N_f=6$ at $J=1.4$ in Fig.~\ref{fig:n6decay} and for $N_f=8$ at $J=2.0$ in Fig.~\ref{fig:n8decay}, respectively.
In Fig.~\ref{fig:n6decay}, the spin-spin and dimer-dimer correlation functions show the similar power-law decay with $2\Delta_S=3.8(2)$ and $2\Delta_D=3.6(3)$.  In Fig.~\ref{fig:n8decay}, both correlation functions decay  with  similar power, with $2\Delta_S=3.8(2)$ and $2\Delta_D=3.4(5)$.
On the whole our data  provides  concrete evidence  that the  deconfined phase in our model  at  various  values of $N_f$  belong to  the algebraic spin liquid~\cite{Rantner2002,Hermele2005,Hermele2007,Xu2007}. One can foresee  that   with a   further increase  of $N_f$, we will reach the expected   power-law behavior $\sim r^{-4}$.

\section{Flux energy per plaquette}
\label{sec:fluxenergy}
\begin{figure}[thp!]
\includegraphics[width=\columnwidth]{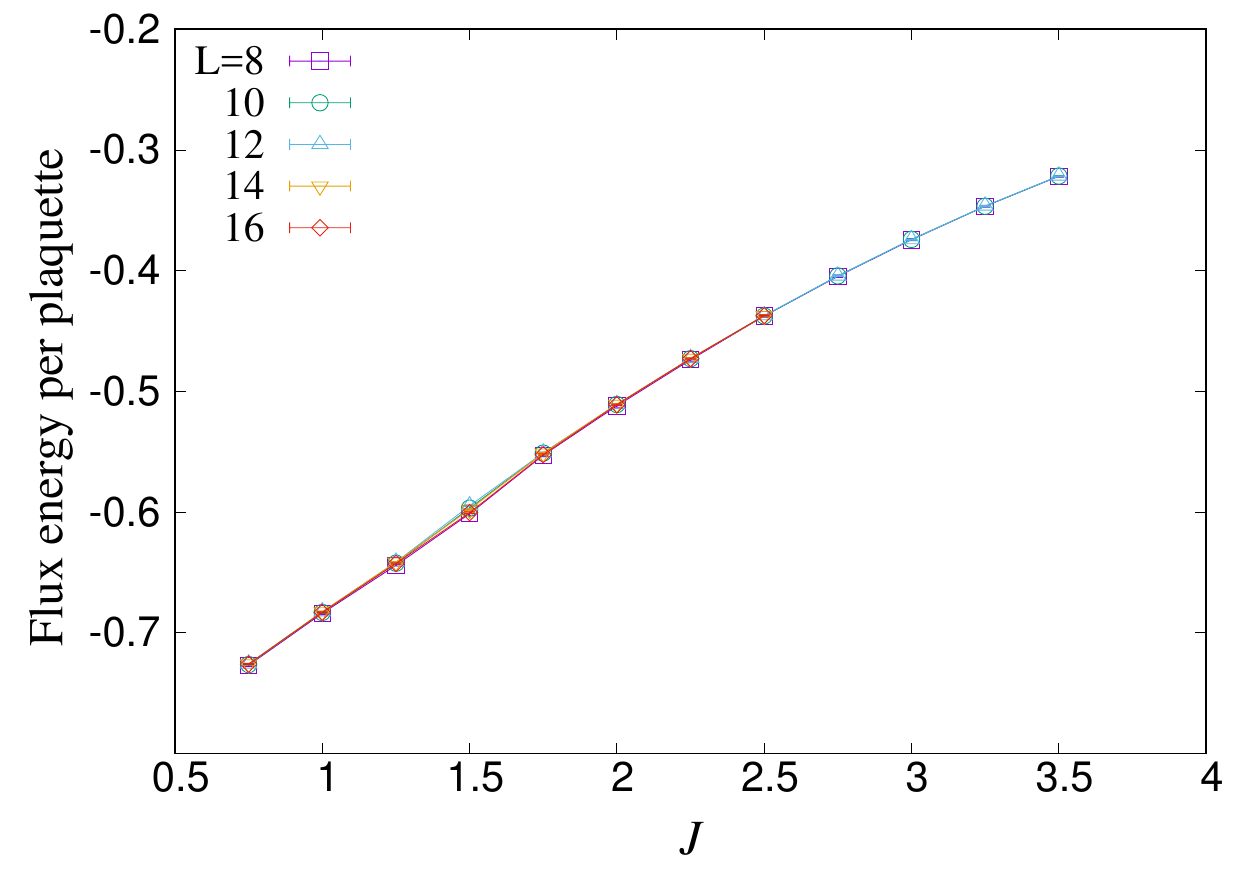}
\caption{Flux energy per plaquette along the same $J$ path. There is no sigularity around $J_c$, suggesting it is a continuous phase transition.}
\label{fig:n2fluxenergy}
\end{figure}
To characterize the continuous nature of confined and deconfined phase transition, we also measured the flux energy per plaquette $\left\langle \frac{1} {L^2} \sum_{\square}\cos \left( \text{curl} \hat{\theta} \right) \right\rangle$. Fig.~\ref{fig:n2fluxenergy}  depicts our result at
$N_f=2$. For other $N_f$'s, the flux energy per plaquette has a similar continuous  behavior.

\begin{figure}[htp]
\includegraphics[width=\columnwidth]{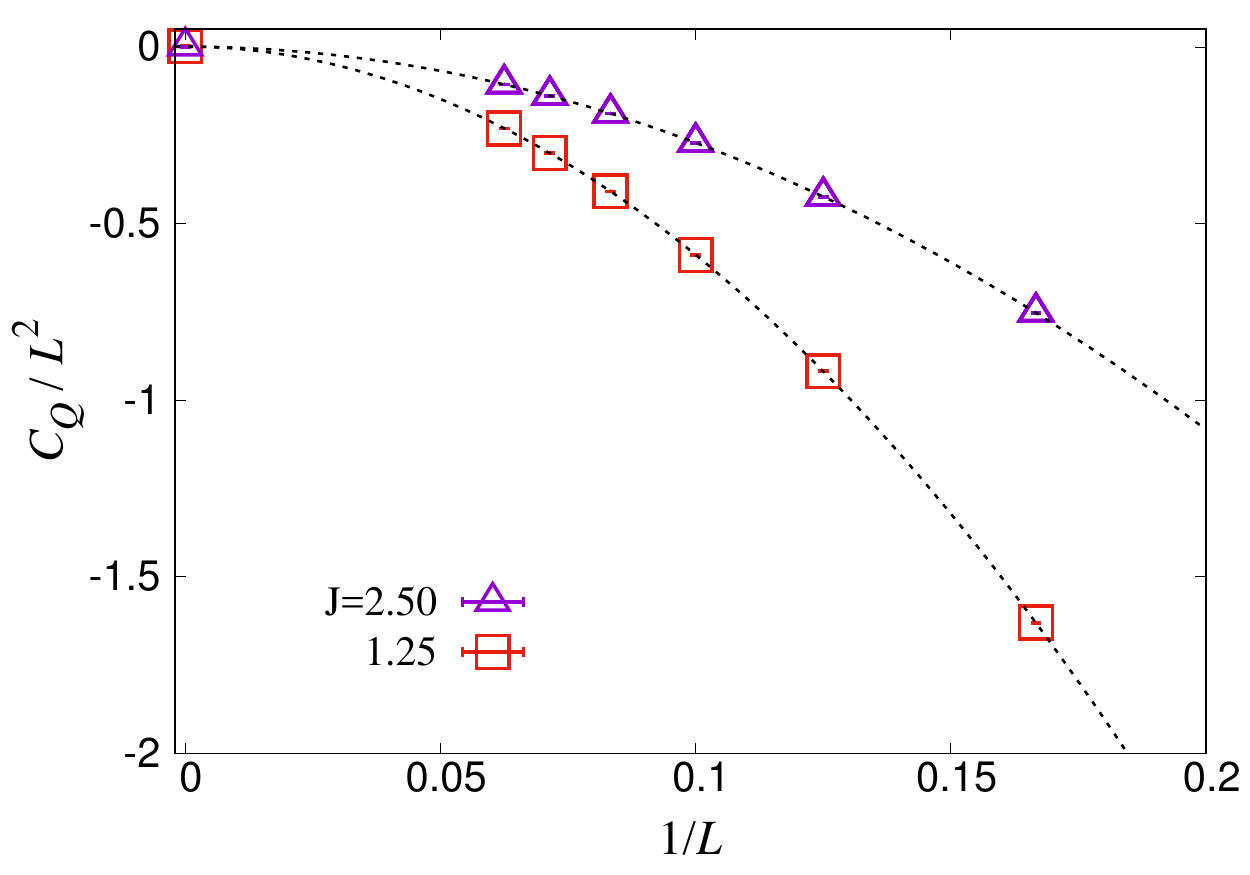}
\caption{Uniform structure factor of $\hat{Q}_i$ (defined in Eq.~\ref{eq:cq}) for $N_f=2$. Note that $\beta=4L$ here. For all $J$s, this uniform structure factor extrapolated to zero in thermodynamic limit. Thus, the constrain of $\hat{Q}_i=0$ is dynamically enforced. }
\label{fig:n2QQ00}
\end{figure}

\section{Dynamically generated constraint}
As mentioned in the  main text, our model corresponds to an unconstrained gauge theory.  As such, the Gauss law  will be dynamically imposed and  $\hat{Q}_i$, defined in Eq.~\ref{eq:Qi},  should converge to  constant  value in the zero temperature limit.  We studied the uniform structure factor of $\hat{Q}_i$ by calculating
\begin{equation}
C_{Q} = \frac{1}{L^2} \sum_{ij} \langle \hat{Q}_i \hat{Q}_j\rangle.
\label{eq:cq}
\end{equation}
We find  that the uniform structure factor of $\hat{Q}_i$ defined above extrapolates to zero in thermodynamic limit as showed in Fig.~\ref{fig:n2QQ00} for $N_f=2$. Other $N_f$ cases show similar behavior.

\begin{figure}[htp]
\includegraphics[width=\columnwidth]{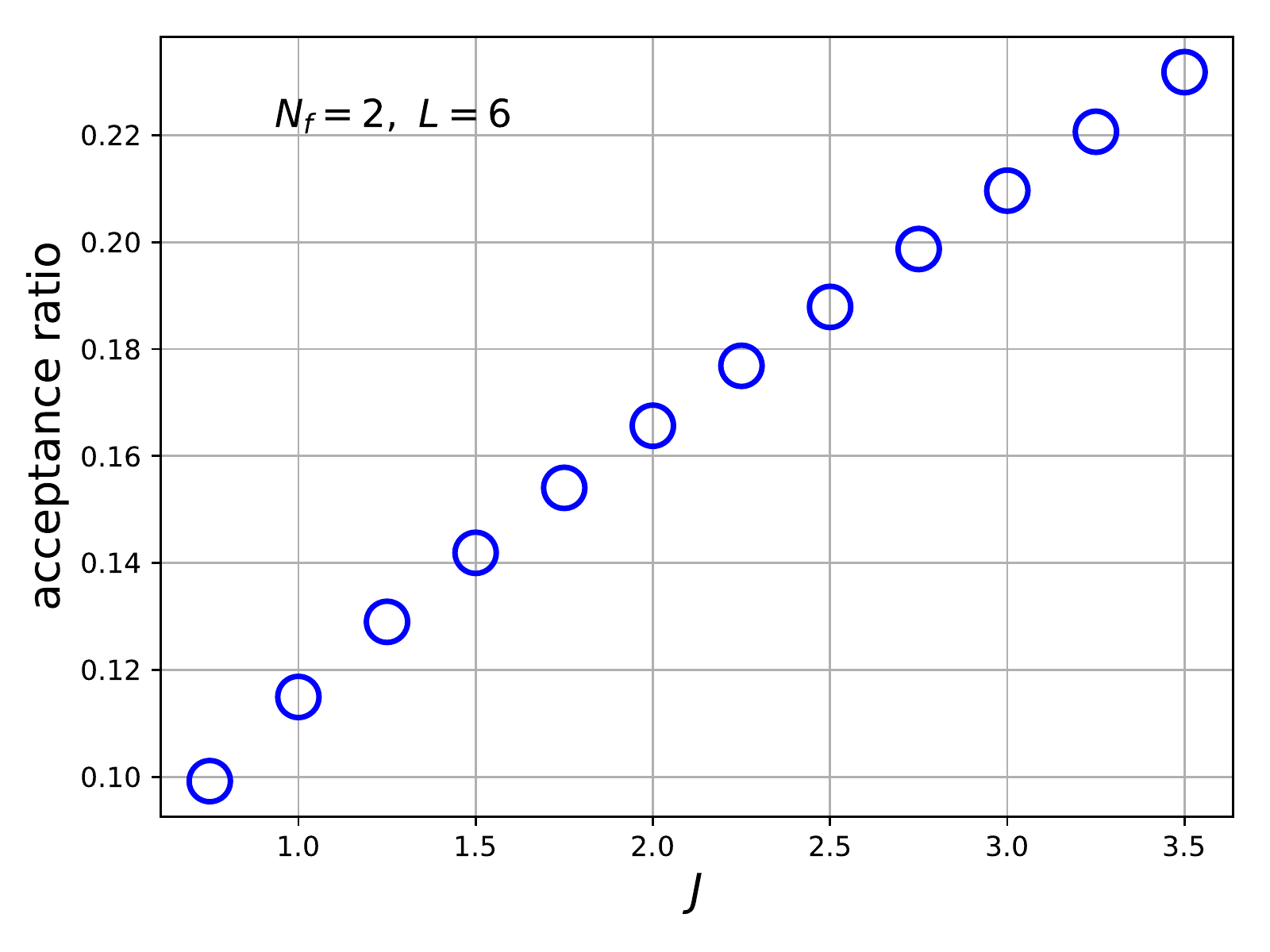}
\caption{Local update acceptance ratio at $N_f=2$ with $L=6$. Note the acceptance ratio for other sizes is almost same, not shown here.}
\label{fig:accep}
\end{figure}

\section{Performance of DQMC on cQED$_3$ coupled to fermionic matter}
As we discussed in the main text, in the DQMC simulation, we use local updates, which flip the gauge variables $\phi_b(\tau)$ ($\in[0,2\pi)$) on space-time lattice one by one and we call one scan of the whole space-time lattice as one sweep, which is usually called one Monte Carlo step in DQMC. For cQED$_3$ problem, we designed a specific fast update method, which greatly improves the computation efficiency, more accurately, by making fast update still work here thus reducing huge time cost for each sweep.

As a first attempt to study this challenging problem of cQED$_3$ coupled to fermionic matter in condensed matter field by DQMC method with local update strategy,
how well does it work here needs a demonstration. Following is a detailed discussion of the performance of the method.

First important quantity associated with the efficiency of the method is the acceptance ratio. Fig.~\ref{fig:accep} illustrated the acceptance ratio for different $J$ at $N_f=2$. The acceptance ratio reduces as $J$ becomes smaller. Fortunately, the acceptance ratio deep in the U1D phase is still quite large, for example at $J=0.75 < Jc=1.6(2)$, the acceptance ratio is $\sim 10\%$.

\begin{figure}[htb]
\includegraphics[width=\columnwidth]{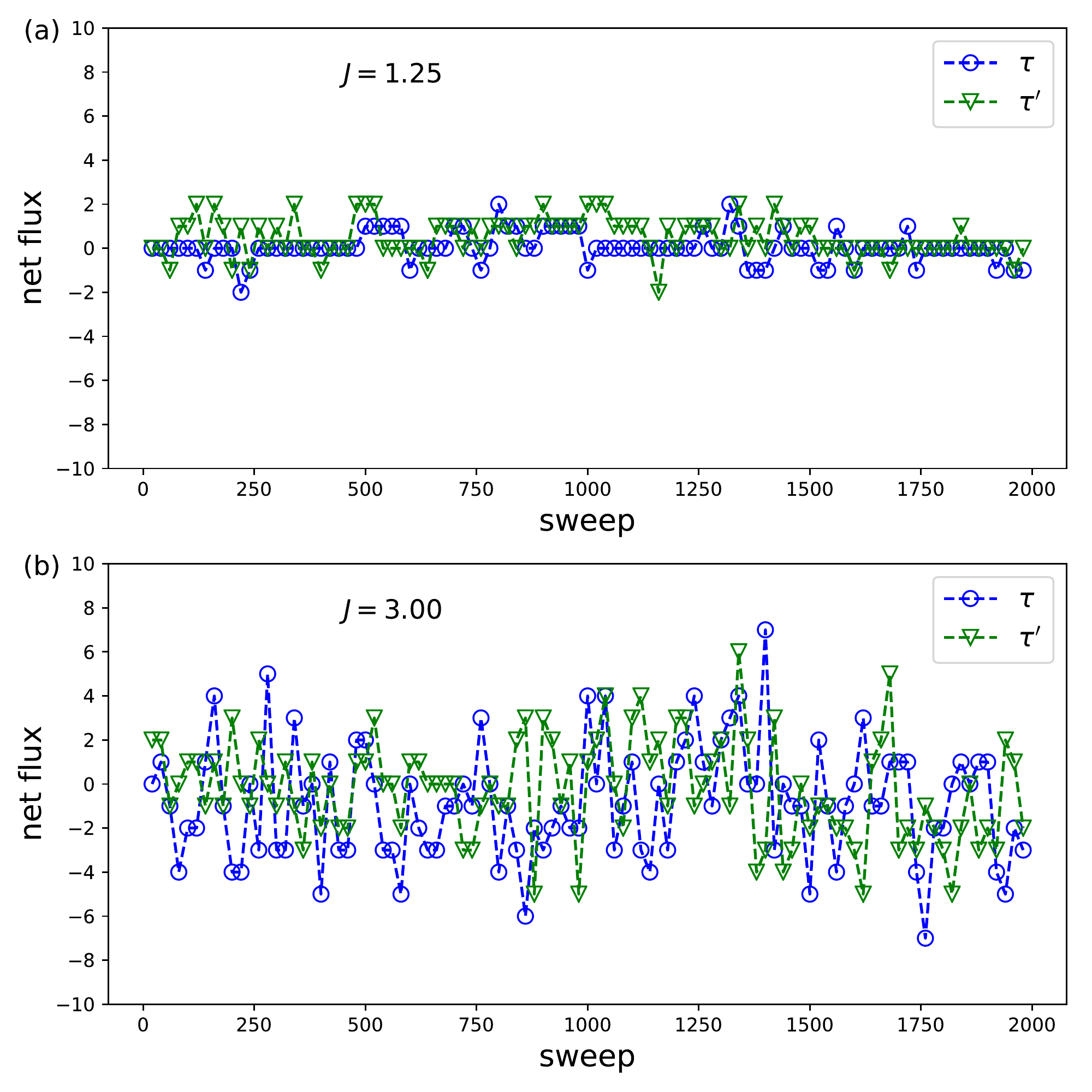}
\caption{Net flux sweep serials in time slice plane $\tau$ and $\tau'$ at $N_f=2$ with $L=12$ (a) inside U1D phase and (b) inside AFM phase. Here $\tau=\Delta\tau$ and $\tau'=8\Delta\tau$. The flux sweep serials are plotted in the interval of 20 sweeps. And a $L^2/2$ is added to shift it to be centred around zero. }
\label{fig:flux_sweep}
\end{figure}
Second important quantity which reflects the efficiency of our method to the specific problem we studied is how quickly does the net flux change in each time plane with Monte Carlo steps.
Flux in each plaquette can be written as $\sum_{b\in \square} \phi_b = \Phi_{\square} + 2\pi m_{\square}$ with $\Phi_{\square} \in [0,2\pi)$ and $m_{\square}$ an integer.
The net flux in one time slice plane $M(\tau)$ is defined as a sum of $m_{\square}$ of each plaquette in the time slice plane $\tau$, $M(\tau)=\sum_{\square} m_{\square}(\tau)$.
Fig.~\ref{fig:flux_sweep} showed such net flux sweep series both inside U1D phase (Fig.~\ref{fig:flux_sweep}(a)) and inside AFM phase (Fig.~\ref{fig:flux_sweep}(b)) at $N_f=2$ with $L=12$ at different time slices $\tau$ and $\tau'$. In the U1D phase, it favors $\pi$($-\pi$)-flux in each plaquette, and the net flux in each time slice plane seldom changes and is weakly fluctuating between different time slices, while in the AFM phase, the net flux changes almost randomly with more extended values and large fluctuations between different time slices plane, as a consequence of proliferate of monopoles.

\bibliography{main_full}

\end{document}